# Electrically Tunable Spin Exchange Splitting in Graphene Hybrid Heterostructure

*Dongwon Shin[1,#], Hyeonbeom Kim[2,#], Sung Ju Hong[3,#], Sehwan Song[4], Yeongju Choi[1], Youngkuk Kim[1], Sungkyun Park[4], Dongseok Suh[2,5,\*], and Woo Seok Choi[1,\*]*

D. Shin, Y. Choi, Y. Kim, W. S. Choi

Department of Physics, Sungkyunkwan University, Suwon 16419, Republic of Korea

E-mail: choiws@skku.edu

H. Kim, D. Suh

Department of Energy Science, Sungkyunkwan University, Suwon 16419, Republic of Korea

E-mail: energy.suh@ewha.ac.kr

S. J. Hong

Division of Science Education, Kangwon National University, Chuncheon 24341, Republic of Korea

S. Song, S. Park

Department of Physics, Pusan National University, Busan 46241, Republic of Korea

D. Suh

Department of Physics, Ewha Womans University, Seoul 03670, Republic of Korea

Keywords: graphene, perovskite $LaCoO_3$, quantum Hall effect, spin exchange splitting, electric gate control, interfacial charge transfer

Graphene, with spin and valley degrees of freedom, fosters unexpected physical and chemical properties for the realization of next-generation quantum devices. However, the spin-symmetry of graphene is rather robustly protected, hampering manipulation of the spin degrees of freedom for the application of spintronic devices such as electric gate-tunable spin filters. We demonstrate that a hybrid heterostructure composed of graphene and $LaCoO_3$ epitaxial thin film exhibits an electrically-tunable spin exchange splitting. The large and adjustable spin exchange splitting of 155.9 – 306.5 meV was obtained by the characteristic shifts in both the spin-symmetry-broken quantum Hall states and the Shubnikov–de Haas oscillations. Strong hybridization induced charge transfer across the hybrid heterointerface has been identified for the observed spin exchange splitting. The substantial and facile controllability of the spin exchange splitting provides an opportunity for spintronics applications with the electrically-tunable spin polarization in hybrid heterostructures.

# 1. Introduction

Symmetry-broken states in graphene, originally with four-fold spin and valley degeneracies, manifest quantum electrodynamic phenomena that are essential for realizing high-speed and dissipation-less quantum devices for spintronics and valleytronics.[1-6] One of the simplest methods to assess degeneracy broken states involves analyzing the quantum Hall effect.[7-9] The quantization effect in graphene is manifested by the quantized Hall conductivity as $\sigma_{xy} = \pm g(N + 1/2)e^2/h$, where $g$ is 4 with both the spin and valley degeneracies, $N$ is a non-negative integer (Landau level; LL), $e$ is the electron charge, $h$ is the Planck's constant, and + and – stand for electron- and hole-dominant carriers, respectively.[10, 11] The filling factor $v$ is defined as $v = \pm g(N + 1/2)$, indicative of LL filling. When the spin and valley degeneracies break under a strong applied magnetic field, $g$ decreases to 2 or 1, leading to $v$ values other than $|v| = 2, 6, 10, \cdots$.[12, 13] Strong exchange interactions are also known to induce spontaneous symmetry breaking of the SU(4) isospin states of graphene.[7-9]

Conventionally, the magnetic proximity effect in heterostructures comprising graphene and magnetic materials has been utilized to introduce exchange interaction to graphene. Whereas doping adatoms or vacancies into graphene have also been applied in an attempt to lift the spin degeneracy,[1, 2, 14-18] these approaches result in random distribution of undesirable defects, which act as scattering centers detrimental to transport properties.[19-21] In contrast, graphene heterostructured with a magnetic layer mostly preserves the intrinsic transport property of graphene. In addition, hybridization between the $\pi$-orbitals in graphene and the spin-polarized $d$- or $f$-orbitals in the adjacent magnetic layer has been suggested to effectively lift the spin degeneracy, despite the sizable van der Waals gap.[19, 22] For example, graphene on magnetic insulators such as yttrium iron garnet, CrSe, EuO, EuS, BiFeO$_3$, or Tm$_3$Fe$_5$O$_{12}$ have shown phenomena such as proximity-induced spin current modulation, strong interfacial exchange field, anomalous Hall effect, and Landau fan shifts[19-25]. On the other hand, if realized, interfacial charge transfer would directly induce spin polarization into graphene. In general, charge transfer has been employed to understand structural and/or electronic reconstruction at the interface.[26, 27] However, exchange spin splitting beyond the magnetic proximity effect could be induced via the direct transfer of spin-polarized charges.[28] We further note that graphene heterostructured with ferromagnetic metal can also exhibit spin exchange splitting with enhanced interfacial charge transfer. However, because of the conduction through the ferromagnetic metallic layer, it would be rather challenging to extract the effect from transport measurements.[29, 30]

**Table 1.** Spin exchange splitting energy $\Delta$ estimated in the graphene/magnetic insulator hybrid heterostructures. A summary of reported $\Delta$ values for graphene/magnetic insulator hybrid heterostructures, depending on the method, system, $T$, and tuning parameters. In the current study, a large and tunable $\Delta$ is observed.[20, 22, 31-35]

| | estimation method | magnetic layer | $T$ [K] | tuning paramter | range | $\Delta$ [meV] | reference |
|---|---|---|---|---|---|---|---|
| calculations | square barrier model | EuO | 0 | temperature | 0~30 K | 0~5 | [31] |
| | the Vienna *ab* initio simulation package | EuO | 0 | interface distance | 2.3~7.8 Å | 0~36 | [32] |
| | | EuO | | the number of magnetic layer | 1~6 | 36~50 | |
| | the Vienna *ab* initio simulation package | EuS | 0 | the number of magnetic layer | 1~6 | 32~40 | [33] |
| | | $Y_3Fe_5O_{12}$ | | | | 64 | |
| | | $CoFe_2O_4$ | | | | 4 | |
| | the Vienna *ab* initio simulation package | $BiFeO_3$ | 0 | interface distance | 2.6~3.0 Å | 70~142 | [34] |
| experiments | non-local measurement | EuS | 4.2 | magnetic field | 0~3.8 T | 0~2 | [20] |
| | Hall measurement & machine learning | CrSe | 2 | field cooling | −9~9 T | 69.4~199.4 | [22] |
| | Hall measurement | $Y_3Fe_5O_{12}$ | 13~300 | temperature | <300 K | <27 | [35] |
| experiments | Hall measurement & fast Fourier transform | $LaCoO_3$ | 2~40 | gate volatage & temperature | −1~1 T & 2~40 K | 155.9~306.5 | current study |

Graphene/LaCoO$_3$ (LCO) hybrid heterostructure is a model system to investigate tunable spin exchange splitting via the interfacial charge transfer. The spin exchange splitting energy $\Delta$ is expressed as $\Delta = cJ_{ex}<S_Z>$, where $c$ denotes the fractional density for ions of $d$ or $f$ electrons to that of itinerant electrons in graphene at the interface, $J_{ex}$ denotes the spatial average of the exchange integral, and $<S_Z>$ denotes the spatial average of spins from the interfacial $d$ or $f$ electrons.[31] Whereas $J_{ex}$ and $<S_Z>$ are constants that are only dependent on the material choice of the magnetic layer, $c$ is determined by the interaction between graphene and the magnetic layer, which can be actively modulated. As summarized in **Table 1**,[20, 22, 31-35] thus far, the experimentally reported values of $\Delta$ lie below 200 meV, with limited tunability.[20, 22, 35] If $\Delta$ can be modulated via the application of a gate electric field, which is one of the most effective tuning parameters of the electrodynamics of graphene, electrically switchable spin filtering devices may be realized.[36] Some of the prerequisites for systematic modulation of $\Delta$ include: (1) high-quality graphene and a magnetic insulating layer with a sharp interface. For example, achieving smoothness is challenging with spinel $CoFe_2O_4$ thin films, hindering high-quality hybrid heterostructure creation. Besides oxide-based materials, 2D van der Waals materials, might also show spin exchange splitting with graphene. However, many van der Waals magnetic materials are unstable in regular environment conditions, limiting practical applications. In addition, large-scale production is challenging, and even methods like chemical vapor deposition yield lower quality samples. (2) robust ferromagnetism with sizable out-of-plane spin components of the magnetic layer to effectively couple with the quantum Hall states of graphene. (3) a *d*-orbital system for the magnetic layer to realize strong *p-d* hybridization with the *p*-orbitals of C. The third point further facilitates charge transfer across the interface. An LCO epitaxial thin film with robust out-of-plane ferromagnetic spin ordering below ~80 K serves as an ideal candidate, with an electronic configuration of $Co^{3+}$ $3d^6$.[37-40] Lastly, (4) oxide-based materials offer versatile functionality and are attainable through precise control in the growth process of oxide thin films.

In this study, we show that the spin-symmetry-broken states in the graphene/LCO hybrid heterostructure exhibit substantial values of $\Delta$ which are largely tunable via the application of a gate electric field and temperature ($T$) modulation. The spin exchange splitting in the graphene/LCO hybrid heterostructure is schematically depicted in **Figure 1**a. The characteristic ferromagnetic ground state of Dirac electrons is demonstrated through quantum Shubnikov–de Haas (SdH) oscillations and quantum Hall plateaus accompanied by spin exchange splitting in the graphene/LCO hybrid heterostructure. We further demonstrate electric modulation of the Fermi energy ($\varepsilon_F$) to modify the up- and down-spin LL energies by introducing a tunable $\Delta$. Such an efficient spin-symmetry breaking in hybrid heterostructure via the interfacial charge transfer paves the way for realizing electrically-tunable spin polarizing devices necessary for next-generation spintronics applications.

## 2. Results and Discussion

Figure 1b shows an optical microscopy image and a schematic of the graphene/LCO hybrid heterostructure device. The hybrid heterostructure constitutes an encapsulating hexagonal boron nitride (h-BN) layer (~15 nm), monolayer graphene, and a 30-nm-thick LCO epitaxial thin film on a $SrTiO_3$ (STO) substrate along the (001) direction, which respectively serve as the top gate dielectric, transport channel, magnetic insulating layer, and substrate, as illustrated in the bottom panel of Figure 1b. The high crystalline quality and highly insulating nature of the epitaxial LCO thin film on the STO substrate was consistently evidenced via X-ray diffraction, X-ray reflectivity, atomic force microscopy, and $T$-dependent resistivity measurements (Figure S1 and S2, Supporting Information).[38, 41, 42] Monolayer graphene was mechanically exfoliated and transferred onto the top surface of the LCO thin film. The quality of the graphene was confirmed via Raman spectroscopy (Figure S3a and S3b, Supporting Information).[43] The homogeneous nature of the hybrid heterostructure was confirmed by the absence of the shift of the G and 2D peaks in the Raman spectroscopy mapping (Figure S3c, Supporting Information). The longitudinal ($R_{xx}$) and transverse ($R_{xy}$) resistances were simultaneously measured with top-gate voltage ($V_g$) modulation through the h-BN layer and the magnetic ($H$) field along the out-of-plane direction of the hybrid heterostructure.

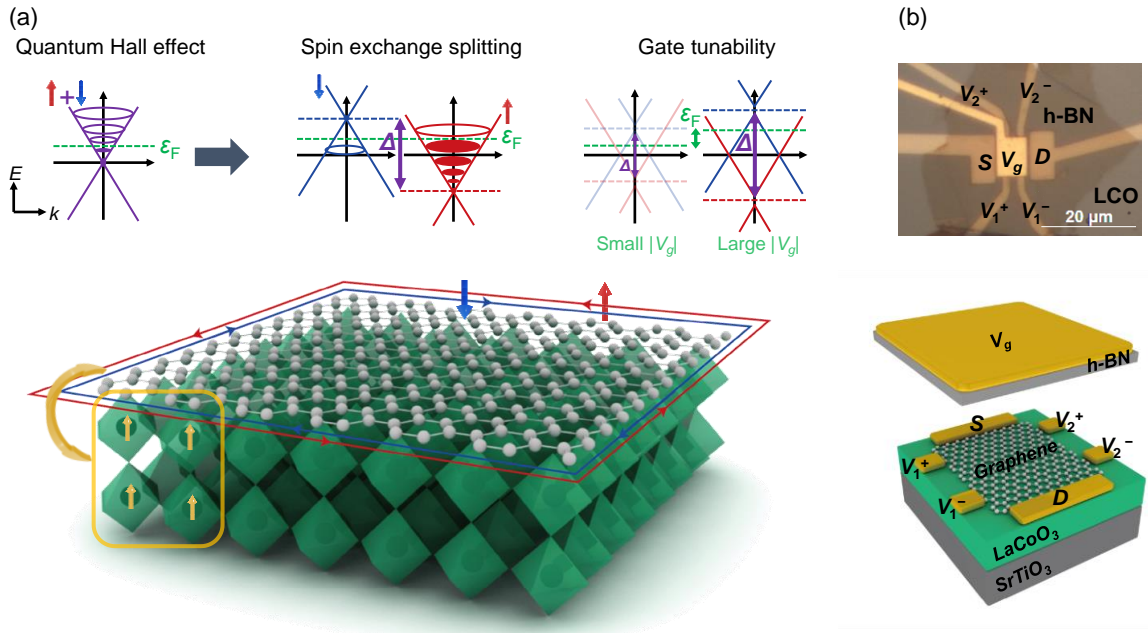

**Figure 1.** Conceptual descriptions for spin-polarized graphene combined with an LCO thin film. a) At the LCO layer (green), octahedrons including Co atoms are shown. Under an external magnetic field, electrons in the interior of the graphene follow closed orbits, and electrons at the edge of graphene form edge states. At a low $T$, the LCO layer exhibits a ferromagnetic phase, introducing spins from Co atoms (yellow arrows) into the adjacent graphene layer via interfacial charge transfer. The exchange interaction from LCO layer effectively split the Dirac cone to up-spin electron-dominant (red-colored scheme) and down-spin hole-dominant (blue-colored scheme) carriers. Up- (red arrow) and down-spins (blue arrow) propagating at the edge of graphene are indicated by red (up-spin) and blue (down-spin) lines, respectively. Furthermore, spin exchange splitting energy can be modulated by gate electric field. b) Optical image and schematic of the graphene/LCO hybrid heterostructure device with an h-BN top gate layer. In the Hall measurements, a current $I$ was applied along $S$ (source) and $D$ (drain), and the longitudinal voltage ($V_{xx}$) was determined between $V_1^+$ ($V_2^+$) and $V_1^-$ ($V_2^-$), and the Hall (transverse) voltage ($V_{xy}$) was determined between $V_1^+$ ($V_1^-$) and $V_2^+$ ($V_2^-$). The longitudinal and Hall resistance are defined as $R_{xx} = V_{xx}/I$ and $R_{xy} = V_{xy}/I$, respectively.

To examine the interlayer interaction across the interface of the hybrid heterostructure, the hybridization between the $\pi$-orbitals in graphene and the spin-polarized $d$-orbitals in the LCO layer is characterized. X-ray magnetic circular dichroism (XMCD) at the C $K$-edge with an out-of-plane magnetic field of 0.8 T shows clear spin polarization below 40 K, as shown in **Figure 2**. The Co $L$-edge spin polarization supports the strong interaction between the C $p$- and Co $d$-orbitals at low $T$, facilitating the charge transfer across the heterointerface (Figure S4a and S4b, Supporting Information). The large enhancement of the spin polarization below 40 K corresponds to the enhancement of ferromagnetic order in the LCO epitaxial thin film, as shown in the $H$-field-dependent magnetization, $M(H)$, curves along the out-of-plane direction (Figure S4c and S4d, Supporting Information). Whereas ferromagnetic spin ordering in conventional LCO epitaxial thin film is predominantly along the in-plane direction with a nominal ferromagnetic $T_C$ of 80 K, the out-of-plane spin ordering with strong enhancement of magnetic coercive field becomes evident below 40 K as shown in our experimental $M(H)$ curve.

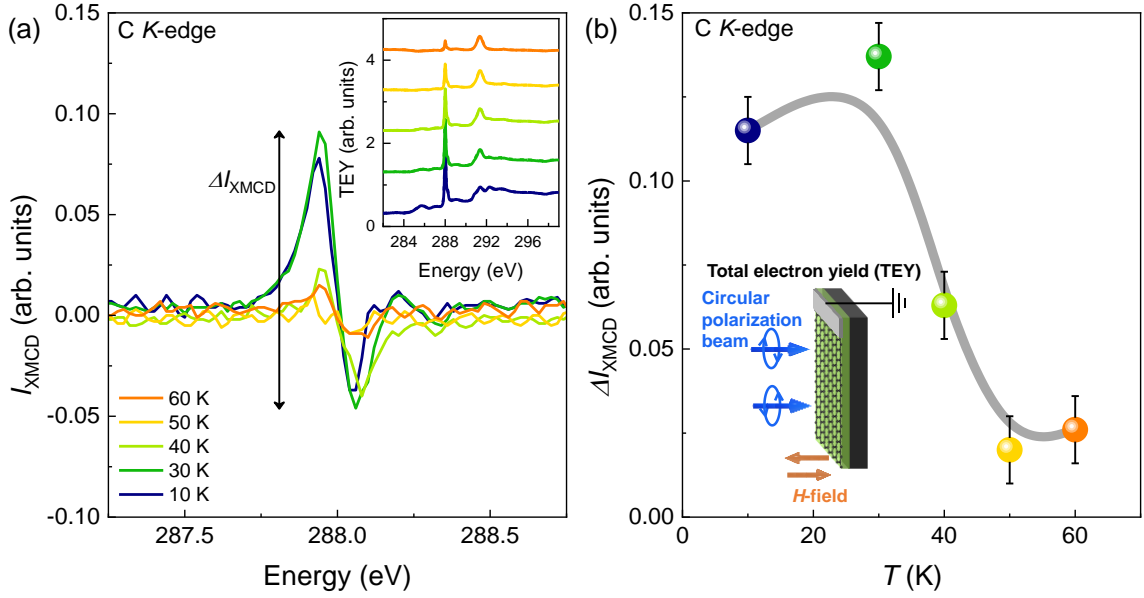

**Figure 2.** Element-specific magnetic characterization in graphene/LCO hybrid heterostructure. a) XMCD asymmetry spectra of C *K*-edge demonstrate clear spin polarization below 40 K at C *K*-edge. The vertical arrow defines $\Delta I_{\text{XMCD}}$ from the peak-to-peak height difference, as summarized in (b). The inset in (a) shows the raw XAS spectra obtained for the C *K*-edge with 0.8 T of the magnetic field along the normal direction of the sample surface. The spectra for right- and left-handed polarizations of the light are not discernible in this range. b) *T*-dependence of $\Delta I_{\text{XMCD}}$ for the C *K*-edge. The inset shows schematics of graphene/LCO hybrid structure with electrode for the XMCD measurement.

Having established the strong interlayer coupling, we characterized the quantized electronic states in the hybrid heterostructure. It is worth noting that the transport properties of graphene/LCO hybrid heterostructure exhibits a typical $V_g$-dependent peak in $R_{xx}$ with a charge neutral point (CNP) under the absence of an *H*-field (Figure S5a, Supporting Information). Considering the $R_{xy}(H)$ at different $V_g$ values at 2 K (Figure S5b, Supporting Information), we categorize Region I and II, wherein quantum Hall plateaus are obscure and evident, respectively (Figure S5b and S5c, Supporting Information). The $R_{xy}(H)$ curves in Region I are obscure because of the simultaneous contribution of both electron and hole carriers. Therefore, we will focus our dicussion on Region II (**Figure 3**). Nevertheless, upon investigating $R_{xx}(V_g)$ near the CNP, i.e., within Region I, we observe evidence of the spin exchange splitting in the nonmonotonic behavior of $R_{xx}$ at the Dirac point (see Section S4 and Figure S6 in Supporting Information).

Figure 3a simultaneously shows the out-of-plane $H$-field-dependent electric resistivities, $R_{xx}(H)$, and $R_{xy}(H)$ at 2 K and $V_g = -1$ V. The quantized states with filling factors $|v| = 6, 10$, and 14 are represented by strong dips in the SdH oscillations ($R_{xx}(H)$) and quantum Hall plateaus ($R_{xy}(H)$; red dashed lines). Interestingly, the spin-symmetry-broken states are observed through the appearance of new quantum Hall states at $|v| = 8$ and 12 as kinks (blue arrows) in $R_{xy}(H)$. Although weak, we emphasize that the new states are consistently observed in repeated experiments on a different graphene/LCO hybrid heterostructure samples (Figure S7a and S7b, Supporting Information). We note that there is no significant dependence on the in-plane $H$-field of graphene itself. Hence, it would be rather difficult to probe the spin exchange splitting with the in-plane $H$-field. Moreover, our transport measurements primarily focus on observing the quantum Hall effect, which necessitates a perpendicular $H$-field. The observed spin-symmetry-broken states ($|v| = 8$ and 12) can be predicted by possible LL energies of both up- and down-spin charge carriers. The possible LL energies at different $|V_g|$ values are schematically illustrated in Figure 3b and 3c, which manifest the possibility of $V_g$ modulation of $\varDelta$. The up- and down-spin LL energies are determined using $E_{N\uparrow\downarrow} = \pm v_F[2e(h/2\pi)|N_{\uparrow\downarrow}|H]^{1/2} \pm \varDelta/2$, where $v_F$ denotes the Fermi velocity ($10^8$ cm/s), and $N_{\uparrow\downarrow}$ denotes the integer LL for the up- and down-spins. $v$, with the spin-symmetry-broken state, can now be estimated as $|v| = |v_\uparrow + v_\downarrow| = |2(N_\uparrow - N_\downarrow)|$ when $\varDelta/2$ is larger than $|\varepsilon_F|$.[11, 22] For example, when a large $|V_g|$ of −1 V is applied (Figure 3b), $|v| = 6, 8, 10, 12, \cdots$ can be expected from the combination of $N_\uparrow = 4, 5, 6, 7, \cdots$ and $N_\downarrow = 1$. On the other hand, when a smaller $|V_g|$ of −0.5 V is applied (Figure 3c), the states come closer together within the same $H$-field range, such that $|v| = 2, 4, 6, 8, 10, 12, \cdots$ can be expected from the combination of $N_\uparrow = 2, 3, 4, 5, 6, 7, \cdots$ and $N_\downarrow = 1$.

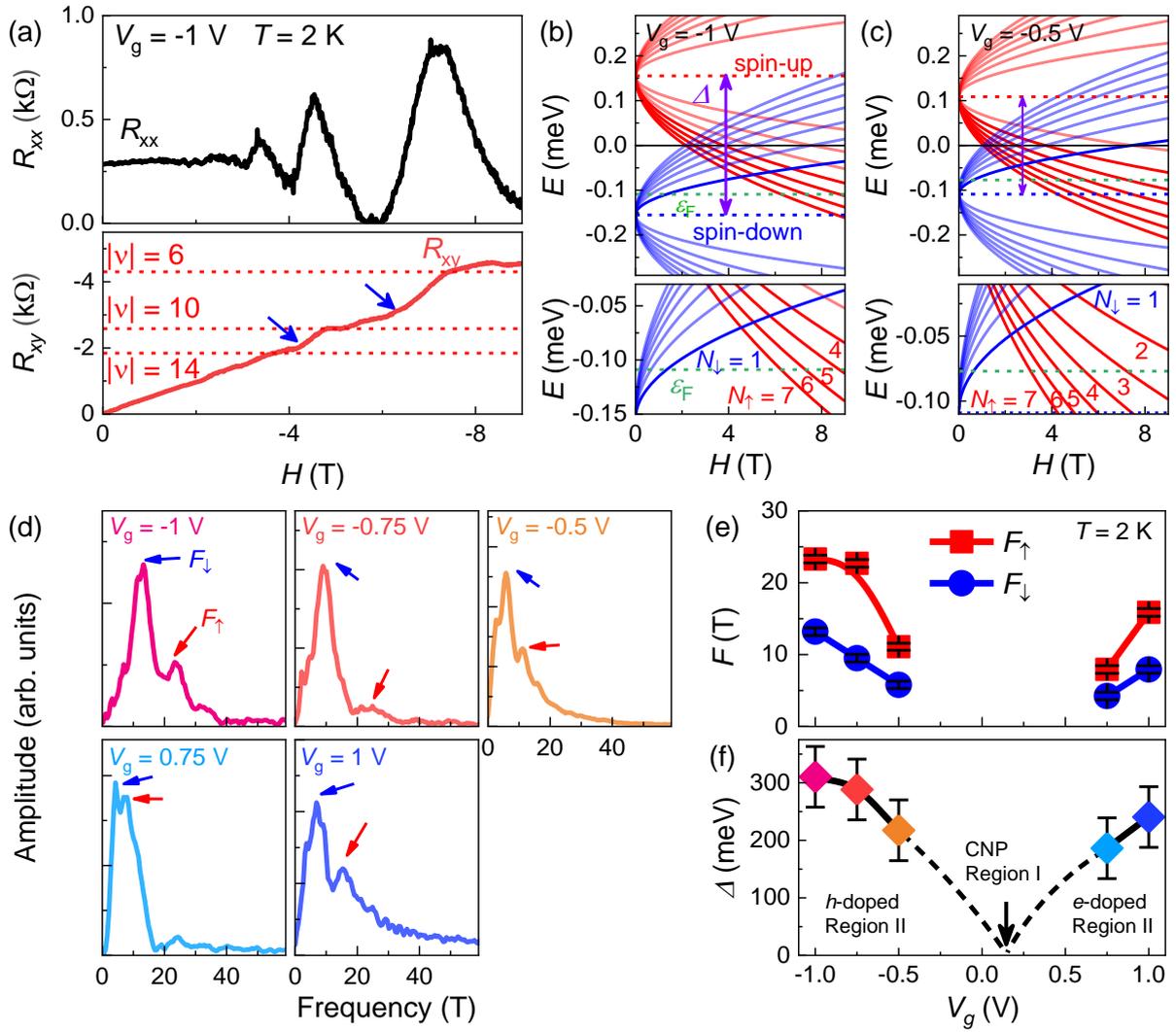

**Figure 3.** Quantized spin states and quantitative calculation of $\Delta$ via the FFT analyses for the graphene/LCO hybrid heterostructure. a) $R_{xx}(H)$ (black curve) and $R_{xy}(H)$ (red curve) for the hybrid heterostructure. Horizontal red-dashed lines indicate conventional quantum Hall states $|\nu| = 6$, 10, and 14. Blue arrows indicate kinks, which demonstrate spin-symmetry broken states such as $|\nu| = 8$ and 12. Schematics of the possible LL energies at b) $V_g = -1$ V (large $|V_g|$) and c) $V_g = -0.5$ V (small $|V_g|$) values with $\Delta$. Red and blue solid lines indicate the up- and down-spin LL energies, respectively. d) $V_g$-dependent FFT analysis with distinct two-peak structures shown as features of up- and down-spin. e) $V_g$-dependent FFT frequency and f) $\Delta$ estimated from d) at 2 K for various $V_g$. The dashed lines are guide to the eye, connecting the experimental data and the CNP in Region I.

The fast Fourier transform (FFT) analysis result further supports the spin exchange splitting in graphene, with gate field tunability, as quantitative $\Delta$ values are characterized. The FFT amplitudes extracted from $R_{xx}(1/H)$ at 2 K for $V_g = -1, -0.75, -0.5, 0.75$, and 1 V are shown in Figure 3d. The peaks in the FFT result indicate SdH oscillation frequencies, $F_\uparrow$ (red arrow) and $F_\downarrow$ (blue arrow), which results are summarized in Figure 3e; thus, the cross-sectional area of the Fermi surfaces for the up- and down-spin charge carriers can be determined according to the Onsager relation. Assuming linear dispersion in graphene, the Fermi surface area can be converted into $\varepsilon_F$ to determine $V_g$-dependent $\Delta$, as shown in Figure 3f. In detail, we determine the $F$ by plotting $R_{xx}$ against $1/H$ and extracting $F$ using FFT analysis. $F$ is calculated using the Onsager relation: $F = (\hbar c/2\pi e)S_F$, where $\hbar$ is $h/2\pi$ (Planck's constant divided by $2\pi$) and $S_F$ is the cross-sectional area of the Fermi surface. The charge carrier density for each band, denoted as $n_{\uparrow\downarrow}$, is found using $n_{\uparrow\downarrow} = g(e/h)F_{\uparrow\downarrow}$, where $n_{\uparrow\downarrow}$ is carrier density, and $F_{\uparrow\downarrow}$ is the frequency of SdH oscillation for the up- and down-spin bands. $\varepsilon_F$ for each band is determined as $\varepsilon_{F\uparrow\downarrow} = \hbar v_F[4\pi(n_{\uparrow\downarrow}/10^{10})/g]^{1/2}$, with $v_F$ as Fermi velocity (approximately $10^8$ cm/s), and the difference in $\varepsilon_F$ between up- and down-spin bands represents the spin exchange splitting energy, $\Delta = |\varepsilon_{F\uparrow}| + |\varepsilon_{F\downarrow}|$ (Section S7 of Supporting Information). Depending on the applied $V_g$, and as $\varepsilon_F$ decreases from a hole-dominant carrier (negative $V_g$) to the CNP and then increases again from the CNP to an electron-dominant carrier (positive $V_g$), $\Delta$ can be widely tuned from 306.5 to 208.0 meV (hole) and from 176.7 to 246.5 meV (electron), respectively. Importantly, the FFT analysis provides information about the Fermi surface of both up- and down spin carriers. Because $\Delta$ determines the relative energy shift between the up- and down-spin bands, it is crucial to obtain both SdH oscillation frequencies $F_\uparrow$ and $F_\downarrow$, especially when considering both can vary systematically depending on $V_g$. The repeated experiments on a different graphene/LCO hybrid heterostructure add further certainty to the systematic behavior of the two-peak structures, supporting the tunable spin exchange splitting (289.7 - 246.2 meV) (Figure S7c and S7d, Supporting Information).

Another strong evidence of the spin exchange splitting in the graphene/LCO hybrid heterostructure is the characteristic $T$-dependence of the quantum transport behavior. $T$-dependent SdH oscillations of $R_{xx}(H)$ at $V_g = -1$ V are shown in **Figure 4**a. In typical graphene, the minima of the SdH oscillation and the quantum Hall pleateau shift toward a higher $H$-field as $T$ increases across the whole temperature regime owing to thermal broadening.[10] The same $T$-dependent trend is observed for the graphene/LCO hybrid heterostructure above 40 K (indicated by the red curved arrow). However, when $T$ is decreased below 40 K, the trend is reversed, that is, the minima of the SdH oscillation shift to a higher $H$-field with decreasing $T$ (indicated by the blue curved arrow). This unconventional *T*-dependent behavior is consistently observed in $R_{xy}(H)$ as well, for the quantum Hall plateaus above and below 40 K (Figure 4b). Such *T*-dependent reversal has been understood by spin splitting. When spin splitting is introduced below 40 K and as it increases further with decreasing *T*, the carrier concentration *n* is expected to increase. This increases the characteristic *H*-field (quantum Hall plateau), and a particular value of *v* can be determined via the relation $v = nh/eH$.[44-46] Therefore, the unconventional *T*-dependence of $R_{xx}(H)$ and $R_{xy}(H)$ below 40 K supports the introduction of *Δ*, as summarized in Figure 4c for different *v* values. Repeated *T*-dependent transport measurements for a different graphene/LCO hybrid heterostructure has also been performed, demonstrating consistent evidence of the same features and trend for $R_{xx}$, $R_{xy}$ and the distinct two-peak structures (Figure S8, Supporting Information). Furthermore, the same trend can be observed for $R_{xx}(H)$ and $R_{xy}(H)$ at different $V_g$ values (= −0.75, −0.5, 0.75, and 1 V) (Figure S9 and S10, Supporting Information). FFT analyses of *T*-dependent $R_{xx}(1/H)$ were also performed. Figure S11 and S12 (hole-dominant charge carriers) and S13 (electron-dominant charge carriers) summarize the FFT results obtained at various *T*s, consistently supporting the existence of two Fermi surfaces below 40 K owing to spin exchange splitting.

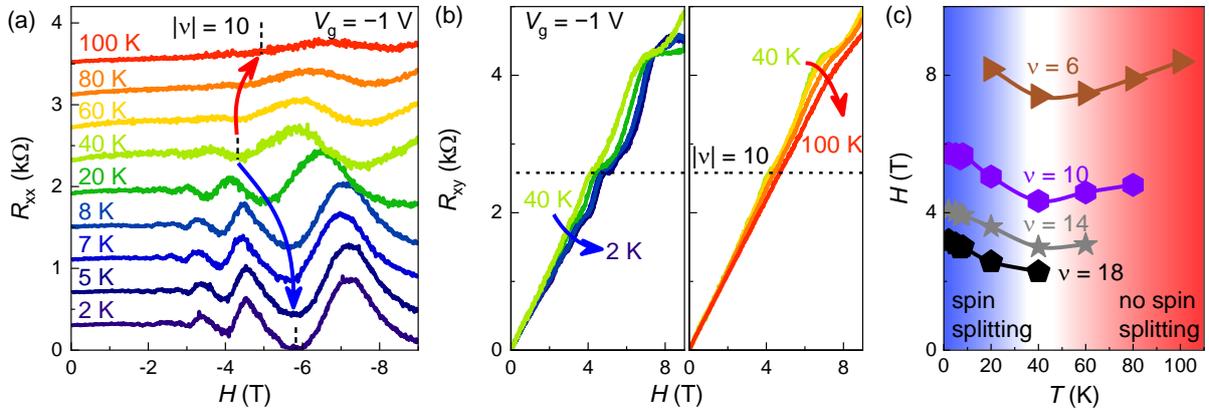

**Figure 4.** Characteristic *T*-dependent behavior of SdH oscillations and quantum Hall plateaus below 40 K. a) $R_{xx}(H)$ at $V_g$ = −1 V at different *T*, vertically shifted (by 0.4 kΩ) for clarify. The vertical dashed lines depict one of the local minima of $R_{xx}$ (|*v*| = 10) which shifts to a higher *H*-field value as the *T* decreases below (blue arrow) or increases above 40 K (red arrow) owing to spin splitting or thermal broadening, respectively. b) Shift of quantum Hall plateaus for $R_{xy}$, revealing the same trend as that in a) below (left panel) and above 40 K (right panel). The black horizontal dashed line indicates |*v*| = 10. c) Summary of *T*-dependent

filling factors ($|v|$ = 6, 10, 14, and 18) at $V_g$ = –1 V. Below and above 40 K are classified as spin splitting and no spin splitting regions.

We do note that the identification of the additional quantum Hall states from the Hall measurements is rather challenging. However, we would like to argue that the graphene/LCO hybrid heterostructure with strong C-Co hybridization indeed exhibits controllable spin exchange splitting, because of the following reasons:

1) The consistent observation of the two-peak structure in the FFT analyses and the systematic change of the peaks as functions of both gate electric field and $T$. By taking advantage of the simple graphene band structure, the most transparent explanation of the two-peak structure is the spin exchange splitting. The systematic movement of the peaks as functions of both gate electric field and $T$ cannot be understood, if the two-peak structures were to originate from extrinsic defects or inhomogeneities.

2) The anomalous $T$-dependent shift of the minima of SdH oscillations, corresponding to the spin ordering $T$ of LCO. The peculiar $T$-dependent shift clearly indicate that our hybrid heterostructure is different from conventional graphene, and is best explained by the spin exchange splitting in graphene.[22]

3) The nonmonotonic behavior of $R_{xx, D}$ with respect to $H$-field also support the spin-symmetry-broken state of graphene. Conventional graphene exhibits a monotonically increasing $R_{xx, D}$ with increasing $H$-field while introducing spin exchange splitting leads to variations in $R_{xx, D}$ with $H$-field due to band crossing and gap opening at edge states.[9, 20]

4) The experimental results are reproducible, with different attempts using different set of hybrid heterostructure samples.

These points collectively support the assertion that our graphene/LCO hybrid heterostructure exhibits controllable spin exchange splitting. Furthermore, we would like to propose a mechanism to elucidate the feasibility of large and tunable spin exchange splitting in the hybrid heterostructure, in addition to the strong hybridization between graphene and LCO. We do expect to observe clearer new quantum Hall states ($|v|$ = 8 and 12) using state-of-the-art growth techniques, such as dry-type transfer and edge-contact metal electrodes. However, we ruled out these methods because they rely on additional plasma treatment, which might damage the magnetic LCO layer.

The experimental $V_g$-tunability of $\Delta$ requires charge transfer across the heterointerface (**Figure 5**a-f), based on our observation of the strong *p-d* hybridization (Figure 2) across the graphene/LCO hybrid heterointerface. To be able to realize the charge transfer across the hybrid heterointerface, the work function difference between the LCO and graphene layers is necessary. Indeed, the work function difference of ~1.2 eV between LCO and graphene (Figure S14, Supporting Information) certifies the development of charge transfer. One of the most important results of our FFT analysis is the enhancement of both up- and down-spin Fermi surface areas, as $|V_g|$ increases (Figure 3e and 3f). This result is highly non-trivial because, for conventional gating, $\varepsilon_F$ would shift within the *identical* band structure of graphene, depending on the electron- (Figure 5b) or hole-doping (Figure 5e). Here, if the Fermi surface area of the up-spin increases, that of the down-spin should decrease. The experimental observation that both the Fermi surface areas increase indicates that the overall shape of the band diagram should be changing (Figure 5c and 5f), necessitating charge transfer. More specifically, when small electron-dominant $V_g$ is applied, a small down-spin hole-dominant Fermi surface (blue circle) and large up-spin electron-dominant Fermi surface (red circle) are formed (Figure 5a). When larger $|V_g|$ is applied, without any charge transfer (Figure 5b), only $\varepsilon_F$ would increase with fixed $\Delta$. This will lead to a smaller down-spin hole-dominant and larger up-spin electron-dominant Fermi surfaces, which is inconsistent with the experimental result. Considering the charge transfer, down-spin electrons from graphene move to LCO layer, creating down-spin holes in graphene (Figure S15a, Supporting Information). As a results, both down-spin hole-dominant and up-spin electron-dominant Fermi surfaces can increase, accompanying the enhanced $\Delta$ (Figure 5c). This is what we observe experimentally (Figure 3e and 3f). A similar analogy can also be applied to hole-dominant $V_g$ (Figure 5d-f and S15b, Supporting Information). A more detailed mechanism considering the interfacial charge transfer and modification of the band structures is shown in Section S8 of Supporting Information (Figure S14 and S15, Supporting Information). By tuning $V_g$, we note a change in $\Delta$ as large as ~50% (from 155.9 to 306.5 meV) in Figure 5g. The value of $\Delta$ = 306.5 meV is the largest value compared to previously reported values, as presented in Table I.

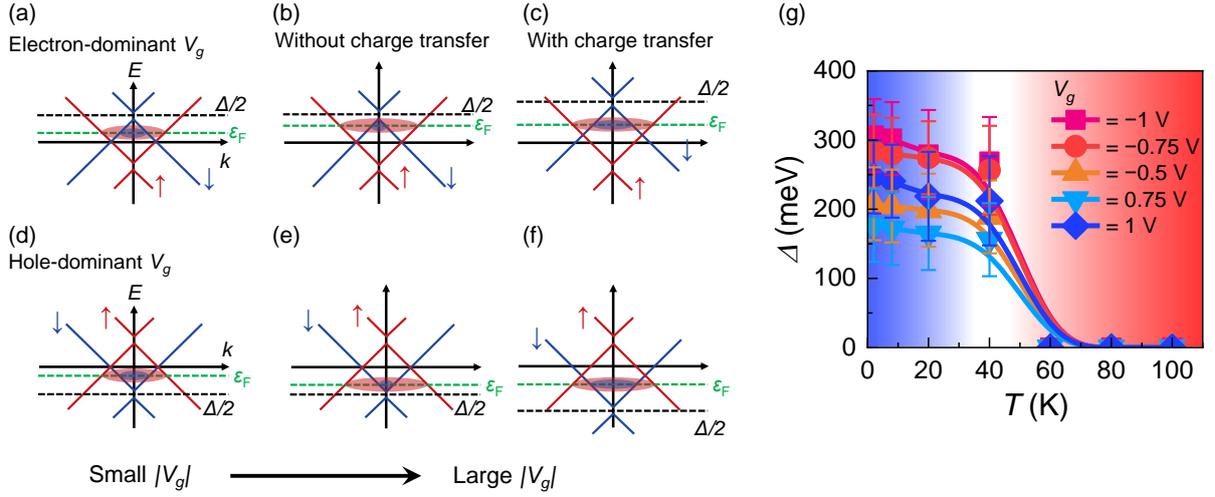

**Figure 5.** Schematic band dispersion in the graphene/LCO hybrid heterostructure via interfacial charge transfer. The red and blue lines represent the linear dispersions corresponding to the up- and down-spins, respectively. Black and green horizontal dotted lines indicate $\Delta/2$ and $\varepsilon_F$, respectively. The band dispersions of small a) electron- and d) hole-dominant $V_g$ are shown at small $|V_g|$. Without the charge transfer, band dispersions of larger b) electron- and e) hole-dominant $V_g$ are identical to a and d, respectively, except for the shift of $\varepsilon_F$ with increasing $|V_g|$. With the charge transfer, band dispersions of c) electron- and f) hole-dominant $V_g$ exhibit enlarged Fermi surfaces for both spins, leading to the larger $\Delta$. g) Summary of $\Delta$ in the graphene/LCO hybrid heterostructure exhibiting substantial $V_g$- and $T$-dependent modulation.

## 3. Conclusion

In conclusion, we demonstrate that the graphene/LCO hybrid heterostructure exhibits an electric gate-tunable spin exchange splitting with energy of $\Delta$ = 306.5 meV via interfacial charge transfer. FFT analyses of the quantum transport behavior represent distinct two-peak structure for the up-and down-spin Fermi surfaces, which enables direct estimation of $\Delta$. A systematic manipulation of $\Delta$ was possible by changing the electric gate voltage and $T$, ranging from 155.9 to 306.5 meV (~ 50%). The characteristic $T$-dependent shift of $v$ below 40 K with the introduction of spin exchange splitting and significant exchange interaction between C $p$-orbital and Co $d$-orbital allows us to consider charge transfer between LCO and graphene, responsible for the large spin exchange splitting. We do note that Rashba-type spin-orbit coupling might play a role owing to the intrinsically broken inversion symmetry of the heterostructure. However, the spin-orbit coupling strength in C and Co is rather small. This effectively limits the spin-orbit gap to several tens of μeV,[1] which is several orders of magnitude smaller than the observed spin exchange splitting. Nevertheless, our experimental observation offers a general method of breaking the spin-symmetry via interfacial charge transfer and envisages electrically switchable spin devices for spintronics applications using graphene.

## 4. Experimental Methods

*Growth of the epitaxial thin film and structural characterizations:* A high-quality LCO epitaxial thin film was fabricated using pulsed laser epitaxy (PLE) on an atomically flat (001)-oriented STO substrate. An excimer KrF laser ($\lambda$ = 248 nm, IPEX864; Lightmachinery) with a fluence of 0.9 J cm$^{-2}$ and a repetition rate of 2 Hz was used. The thin film was synthesized at 500 °C under an oxygen partial pressure of 100 mTorr. The surface of the LCO thin film was examined via atomic force microscopy (AFM; NX10, Park Systems). The atomic structure and crystallinity of the epitaxial thin film were characterized using high-resolution X-ray diffraction (XRD; PANalytical X'Pert Pro). The XRD reciprocal space maps confirmed that the thin film was fully strained. The thickness of the thin film was determined to be 30 nm from the X-ray reflectivity.

*Exfoliation and characterization of a monolayer graphene flake*: A monolayer graphene flake was exfoliated from highly ordered pyrolytic graphite on polyvinyl alcohol/polymethyl methacrylate (PMMA 950 K C4, Microchem) coated onto a SiO$_2$/Si substrate. Raman spectroscopy (Alpha-300s, WITec Instrument GmbH) was used to examine the monolayer graphene with an excitation laser ($\lambda$ = 532 nm) under ambient conditions. The G (~1,580 cm$^{-1}$) and 2D (~2,700 cm$^{-1}$) peaks demonstrated a small full width at half maximum (< 30 cm$^{-1}$). The absence of the D (~1,350 cm$^{-1}$) peak indicated the high quality of the flake with a low level of defects. AFM (AFM5100N, Hitachi Hightech) was used to characterize the thickness of the transferred h-BN (~15 nm) layer on the graphene device.

*Transfer and patterning of graphene on the LCO epitaxial thin film*: A monolayer graphene flake was transferred onto the LCO epitaxial thin film via a typical transfer method. After transferring the graphene flake, the PMMA residue was removed with acetone and rinsed with isopropyl alcohol. A Hall bar structure was fabricated using electron-beam lithography (EBL), and a Cr/Au (5 nm/50 nm) electrode was deposited using a thermal evaporation method. A 15-nm-thick h-BN (HQ Graphene) flake was used as the top dielectric material, covering the graphene device with a polydimethylsiloxane (PF-40/17-X5, Gel-Pac) layer.[47] Another Cr/Au (5 nm/80 nm) top electrode was fabricated using EBL and thermal evaporation.

*X-ray magnetic circular dichroism (XMCD) measurements*: C *K*-edge and Co *L*-edge XMCD measurements were performed in the 6A beamline of Pohang Accelerator Laboratory in the total electron yield mode within a *T* range of 10–60 K under a magnetic field strength of 0.8 T. A circularly polarized beam and external magnetic field were applied in a direction normal to the sample surface.

*Magnetization measurements*: A magnetic property measurement system (MPMS3; Quantum Design) was used to characterize the out-of-plane magnetic properties of the thin films. The magnetic-field-dependent magnetization $M(H)$ was measured in a range of −5 to 5 T at various *T*s between 2 and 100 K.

*Electrical transport measurements*: Longitudinal ($R_{xx}$) and transverse ($R_{xy}$) resistances were simultaneously measured via top-gate modulation through h-BN with a six-terminal Hall bar geometry. Electrical measurements were performed using Keithley 4200A-SCS within a *T* range of 2–300 K and a magnetic field of up to 9 T in a physical property measurement system (PPMS 9T, Quantum Design). High resistance of over 14 GΩ for the LCO thin film under 100 K guaranteed that the electrical current only flowed through the graphene channel. The carrier concentration was effectively tuned with $V_g$.

**Supporting Information**

Supporting Information is available from the Wiley Online Library or from the author.

**Acknowledgements**

Dongwon Shin, Hyeonbeom Kim and Sung Ju Hong contributed equally to this work. The authors thank S. Hong in Sejong University and J. Kim in Incheon National University, Republic of Korea for helpful discussion. This work was supported by the Basic Science Research Programs through the National Research Foundation of Korea (NRF-2021R1A2C2011340, NRF-RS-2023-00220471, NRF-2022M3I7A3051578, NRF-2019R1I1A01058123, NRF-2022H1D3A3A01077468, NRF-2020M3H4A2084417).

**Conflict of Interest**

The authors declare no conflict of interest.

[39] E.-J. Guo, R. Desautels, D. Keavney, M. A. Roldan, B. J. Kirby, D. Lee, Z. Liao, T. Charlton, A. Herklotz, T. Z. Ward, M. R. Fitzsimmons, H. N. Lee, *Science Advanced* **2019**, *5* eaav5050.

[40] D. Shin, S. Yoon, S. Song, S. Park, H. N. Lee, W. S. Choi, *Adv. Mater. Interfaces* **2022**, *9*, 2200433.

[41] D. Fuchs, C. Pinta, T. Schwarz, P. Schweiss, P. Nagel, S. Schuppler, R. Schneider, M. Merz, G. Roth, H. v. Löhneysen, *Phys. Rev. B* **2007**, *75*, 144402.

[42] Y. Li, S. J. Peng, D. J. Wang, K. M. Wu, S. H. Wang, *AIP Adv.* **2018**, *8*, 056317.

[43] Y. y. Wang, Z. h. Ni, T. Yu, Z. X. Shen, H. m. Wang, Y. h. Wu, W. Chen, A. T. S. Wee, *J. Phys. Chem. C* **2008**, *112*, 10637.

[44] J.-M. Poumirol, W. Escoffier, A. Kumar, B. Raquet, M. Goiran, *Phys. Rev. B* **2010**, *82*, 121401(R).

[45] B. E. Feldman, B. Krauss, J. H. Smet, A. Yacoby, *Science* **2012**, *337*, 1196.

[46] T. K. Chau, D. Suh, H. Kang, *Current Applied Physics* **2021**, *23*, 26.

[47] R. Frisenda, E. Navarro-Moratalla, P. Gant, D. Perez De Lara, P. Jarillo-Herrero, R. V. Gorbachev, A. Castellanos-Gomez, *Chem. Soc. Rev.* **2018**, *47*, 53.

**A substantial and tunable spin exchange splitting energy is observed in graphene/LaCoO$_3$ hybrid heterostructure via electrical gate control. Fast Fourier transform analysis enables the determination of the spin exchange splitting energy based on distinct two-peak structures originating from up- and down-spin bands. Interfacial charge transfer across the heterointerface is attributed for the electrically-tunable spin exchange splitting.**

D. Shin[#], H. Kim[#], S. J. Hong[#], S. Song, Y. Choi, Y. Kim, S. Park, D. Suh[*], W. S. Choi[*]

**Electrically-Tunable Spin Exchange Splitting in Graphene Hybrid Heterostructure**

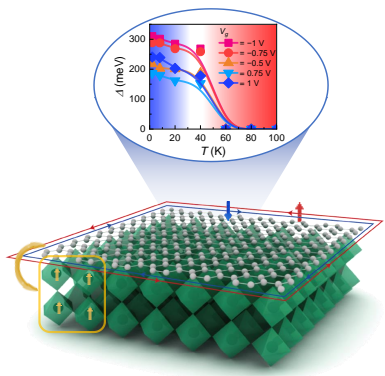

**Electrically Tunable Spin Exchange Splitting in Graphene Hybrid Heterostructure**

*Dongwon Shin[1,#], Hyeonbeom Kim[2,#], Sung Ju Hong[3,#], Sehwan Song[4], Yeongju Choi[1], Youngkuk Kim[1], Sungkyun Park[4], Dongseok Suh[2,5,\*], and Woo Seok Choi[1,\*]*

[1]Department of Physics, Sungkyunkwan University, Suwon 16419, Republic of Korea

[2]Department of Energy Science, Sungkyunkwan University, Suwon 16419, Republic of Korea

[3]Division of Science Education, Kangwon National University, Chuncheon 24341, Republic of Korea

[4]Department of Physics, Pusan National University, Busan 46241, Republic of Korea

[5]Department of Physics, Ewha Womans University, Seoul 03670, Republic of Korea

[\*]Corresponding authors: energy.suh@ewha.ac.kr, choiws@skku.edu

# S1. Fabrication of graphene/LCO hybrid heterostructure

To fabricate graphene/LCO hybrid heterostructure, atomically flat surface of LCO thin film on STO substrate and high-quality graphene can guarantee the sufficient attachment between both materials. In the well-constructed graphene/LCO hybrid heterostructure, the LCO layer can effectively induce spins into the graphene layer.

## S1.1. Structural characterization of the epitaxial LCO thin film

The high-quality LCO thin film was epitaxially grown on the STO (001) substrate using PLE. The additional peaks except for peaks from LCO and STO are not observed in the range of 10 to 80° for $2\theta$ in Figure S1a. Atomically flat surfaces of STO substrate and LCO thin film are shown in the inset of **Figure S1**a. The surface roughness of LCO thin film is estimated to be less than 0.4 nm (1 unit cell of LCO thin film). Figure S1b shows LCO thin film on STO substrate of XRD $\theta$-$2\theta$ scans near the (002) Bragg plane of the epitaxial LCO thin film (#) on STO substrate (*). The 30 nm-thickness of LCO thin film is estimated by XRR measurement in Figure S1c. Figure S1d shows the XRD reciprocal space maps (RSM) around the (103) Bragg reflections for the fully strained LCO thin film grown on STO substrate with tensile strain.

## S1.2. The transport property of LCO thin film on STO substrate

$T$-dependent in-plane electric transport property of LCO thin film on STO substrate is shown in **Figure S2**. Clear insulating behavior is shown for the film, in which the resistance reaches over the experimental limit ($R_{xx}$ > 15 GΩ) below 100 K. Since LCO thin film conduction can be negligible, only the graphene conduction can be determined precisely from the heterostructure. Thus, LCO thin film is only responsible for introducing spins into the graphene layer.

## S1.3. Confirmation of the high-quality monolayer graphene

A monolayer graphene flake was prepared by mechanical exfoliation on Si/SiO$_2$ substrate and was confirmed by Raman spectroscopy measurement. The mechanically exfoliated graphene flake was transferred on top of LCO thin film via a typical transfer method. 15 μm-width of monolayer graphene is placed as black dashed lines in **Figure S3**a. A monolayer graphene shows clear G and 2D peaks fitted by a single Lorentzian curve in Figure S3b. The homogeneous feature of hybrid heterostructure is also proven by the absence of the shift of the G and 2D peak positions from Raman spectroscopy mapping as shown in Figure S3c.

## S2. XMCD measurements and the magnetic properties of graphene/LCO hybrid heterostructure

### S2.1. XMCD measurements along the out-of-plane directions

To further support the spin splitting in graphene/LCO hybrid heterostructure below 40 K, we additionally conducted the $T$-dependent XMCD measurement at Co $L$- and C $K$-edge. At low $T$, the Co atom from the LCO layer has spins, which can be detected at Co $L$-edge (770 – 810 eV). However, the C atom initially has no spin, XMCD result at C $K$-edge (287 – 289 eV) is not shown with external magnetic field. If graphene/LCO hybrid heterostructure is successfully fabricated, spins from Co atom can be induced into graphene layer to show spin signal at C $K$-edge. The XMCD asymmetry spectra was calculated with the average of the differences between two different polarizations (left- and right-handed) for each magnetic field direction. $I_{XMCD} = [(TEY_{+H, left} − TEY_{+H, right}) + (TEY_{−H, right} − TEY_{−H, left})]/2$, where $+H/−H$ and left/right stand for external magnetic field direction and handedness of the circular polarization, respectively. **Figure S4**a and S4b show XAS and XMCD spectrum obtained near Co $L$-edge below 60 K for both right and left circular polarizations. Induced spin signal at C $K$-edge is also shown up to 40 K, and is abruptly suppressed above 40 K, as shown in Figure 2. It is worth noting that the spin splitting in graphene along the out-of-plane direction could be induced by LCO layer below 40 K.

### S2.2. MPMS measurements along the out-of-plane direction

The magnetic field dependent magnetization, $M(H)$, measured up to 5 T with various $T$s (2, 10, 20, 40, 60, 80, and 100 K), demonstrates clear ferromagnetic properties, as shown in Figure S4c. In particular, we found a hysteresis loop opening below 40 K, indicating evidence of out-of-plane spin ordering. Figure S4d exhibits the coercive field as a function of $T$, clearly supporting the ferromagnetism below 40 K.

## S3. Electrical transport measurements in graphene/LCO hybrid heterostructure

**Figure S5**a shows $R_{xx}(V_g)$ at 2 K, which exhibits a CNP at ~0.15 V, similar to that reported for conventional monolayer graphene with slight hole-doping.[1] Our hybrid heterostructure exhibits charge mobility of > 80,000 cm$^2$ V$^{-1}$ s$^{-1}$, which is comparable with that of conventional graphene on SiO$_2$/Si substrates, indicating high structural and electronic quality.[2] $R_{xx}(V_g)$ can be classified into regions with relatively low (Region I, small $|V_g|$) and high carrier densities (Region II, large $|V_g|$), depending on $V_g$. Region I is characterized as the vicinity of the CNP, wherein the hole and electron charge carriers coexist (yellow and green regions) and evenly contribute to conduction. Region II is characterized as the region away from the CNP, wherein only one type of charge carriers, that is, either electrons (blue region, large positive $V_g$) or holes (red region, large negative $V_g$), dominates conduction. Figure S5 and S5c show $H$-field-dependent $R_{xy}$, $R_{xy}(H)$, at different fixed $V_g$ values for Region I and Region II, respectively. In Region I, the mixed contributions from both electron- and hole-dominant carriers obscure the quantized analyses. Similar distorted quantum Hall plateaus have been observed for graphene on the SiO$_2$ and GaAs substrates near the CNP.[3, 4]

## S4. The signature of spin exchange splitting near CNP

The significant spin exchange splitting is observed in the $V_g$-dependent $R_{xx}$ curves across various $H$-fields (**Figure S6**a). A close examination of the $R_{xx}$ maxima in graphene/LCO hybrid heterostructure reveals the nonmonotonic behavior of $R_{xx}$ at the Dirac point ($R_{xx, D}$) in Figure S6b, in contrast to conventional graphene, where $R_{xx, D}$ monotonically increases with the applied $H$-field.[5, 6] This nonmonotonic trend of $R_{xx, D}$ for graphene-based hybrid heterostructures has been reported in previous results, and interpreted as an evidence of the spin exchange splitting.[5, 6] We can explain the nonmonotonic change in $R_{xx, D}$ in different $H$-field regimes using graphene band structure splitting. In the range of $0 < H \leq 1$ T, the increase in $R_{xx, D}$ is typically attributed to the presence of a perpendicular external $H$-field. The decrease of $R_{xx, D}$ in the range of $1 < H \leq 2$ T suggests the presence of a spin-polarized state. The up-and down-band crossing at the edge state can create a conducting path, resulting in reduction in $R_{xx, D}$. Interestingly, we observe an increase in $R_{xx, D}$ on increasing $H$-field in the high $H$-field regime above 2 T. This behavior indicates the opening of a the gap between up- and down-spin band in the edge excitation spectrum. However, we do note that this $H$-field dependent change in the gap is about two orders of magnitude smaller than the spin exchange splitting caused by the LCO layer.

## S5. Additional electric transport measurements for the graphene/LCO hybrid heterostructure

Since each transferred hybrid heterostructure has different physical contact, there is heterostructure-to-heterostructure variation in the coupling strength which might lead to different results. However, we have conducted additional experiments for a different set of the hybrid heterostructure. Experimental results including the $V_g$- and $T$-dependent spin exchange splitting were consistently shown. In particular, not only that new quantized states ($|v|$ = 8 and 12) are more clearly visible, but also distinct two-peak structures are shown. These results were inserted in the Supplementary Information of **Figure S7** and **S8**.

## S6. The shift of SdH oscillations and quantum Hall plateaus in graphene/LCO heterostructure

Additional $T$-dependent transport measurements were performed at $V_g$ = −0.75, −0.5, 0.75, and 1 V for the hybrid heterostructure. **Figure S9** and **S10** show $R_{xx}$, $R_{xy}$, and obtained filling factors with respect to the $T$ for hole and electron type charge carriers, respectively, in Region II. Same as the case of $V_g$ = −1 V (Figure 4), graphene/LCO hybrid heterostructure exhibits consistent $T$-dependence. The result consistently supports the existence of spin splitting in Region II.

## S7. Estimation for the cross-sectional area of the Fermi surface and $\Delta$ via FFT analysis

**Figure S11**a shows the FFT of the SdH oscillations extracted at 2, 20, 40, 60, 80, and 100 K by $R_{xx}(1/H)$. Figure S11a at 2 K demonstrates two distinct peaks $F_\uparrow$ (red arrow) and $F_\downarrow$ (blue arrow) are shown up to 40 K (upper panels of Figure S11a). Whereas a clear spin splitting is obtained at and below 40 K, only one peak $F$ (purple arrow) is shown above 40 K (bottom panels of Figure S11a). Extracted $F$ from 2 to 100 K at −1 V is summarized in Figure S11b. Same trends for the hole- (**Figure S12**) and electron- (**Figure S13**) dominant carriers are observed consistently.

**S7.1. Onsager relation to determining the frequency of SdH oscillation *F***

The Fermi surface is composed of two spin-degenerated pockets (up- and down-spin) in graphene. When spin degeneracy is broken by exchange interaction, Fermi surface can be split into up- and down-spin pockets, resulting in the existence of two Fermi surfaces. Using FFT analysis, plotted $R_{xx}(1/H)$ is transformed to $F$ (the frequency of SdH oscillation) dependence on amplitude. $F$ is expressed by the Onsager relation, which is defined as, $F = (\hbar c/2\pi e)S_F$, where $\hbar$ is $h/2\pi$ ($h$ is Plank's constant), $c$ is the light velocity, $e$ is the electron charge and $S_F$ is the cross-sectional area of the Fermi surface.[7]

**S7.2. Calculation of *Δ***

Using $F$, the charge carrier density of each band can be determined as $n_{\uparrow\downarrow} = g(e/h)F_{\uparrow\downarrow}$, where $n_{\uparrow\downarrow}$ is carrier density, $g$ is degeneracy, $e$ is the electron charge, $h$ is the Plank's constant, and $F_{\uparrow\downarrow}$ is the frequency of SdH oscillation.[8] According to $n_{\uparrow\downarrow}$, Fermi energy $\varepsilon_F$ is also described by $\varepsilon_{F\uparrow\downarrow} = \hbar v_F[4\pi(n_{\uparrow\downarrow}/10^{10})/g]^{1/2}$, where $\hbar$ is $h/2\pi$ ($h$ is Plank's constant), $v_F$ is the Fermi velocity ($10^8$ cm/s), $n_{\uparrow\downarrow}$ is carrier density of up- and down-spin band and $g$ is degeneracy.[9] Determination of $\varepsilon_{F\uparrow\downarrow}$ difference between up- and down-spin band could be considered as spin exchange splitting energy $\Delta = |\varepsilon_{F\uparrow}| + |\varepsilon_{F\downarrow}|$.

## S8. Interfacial charge transfer in the graphene/LCO heterostructure by electric-field gating

Epitaxial LCO thin film on STO substrate exhibits the high spin $Co^{3+}$ $3d^6$ configuration, with four $t_{2g}$ electrons (three up-spins and one down-spin) and two $e_g$ electrons (two up-spins). Based on the high spin configuration of LCO, only down-spin electrons can be transferred. Considering the work function of graphene (4.57 eV) and LCO (5.73 eV),[10, 11] intrinsic charge transfer would occur with electrons moving from graphene toward LCO layer (**Figure S14**). When electron-dominant $V_g$ is applied (**Figure S15**a), $\varepsilon_F$ would shift upward. This would increase in up-spin electrons (red dotted arrows) and decrease in down-spin holes (blue dotted arrows) without any charge transfer. Spins in black circles indicate spin configuration without the charge transfer. When charge transfer occurs, down-spin electrons in graphene would transfer to LCO, creating down-spin hole states (additional spins in orange circles). Such a trend would be enhanced upon increasing $|V_g|$. On the other hand, when hole-dominant $V_g$ is applied (Figure S15b), $\varepsilon_F$ would shift downward. This would increase in up-spin holes (red dotted arrows) and decrease in down-spin electrons (blue solid arrows) without any charge transfer. When charge transfer occurs, less down spin electrons would transfer to LCO upon increasing $|V_g|$. This would effectively increase the down spin electron states in graphene. This charge transfer mechanism can effectively explain the experimentally observed increasing $\varDelta$ and Fermi surface frequency of both up-and down-spins with $|V_g|$ increase (Figure 3e and 3f).

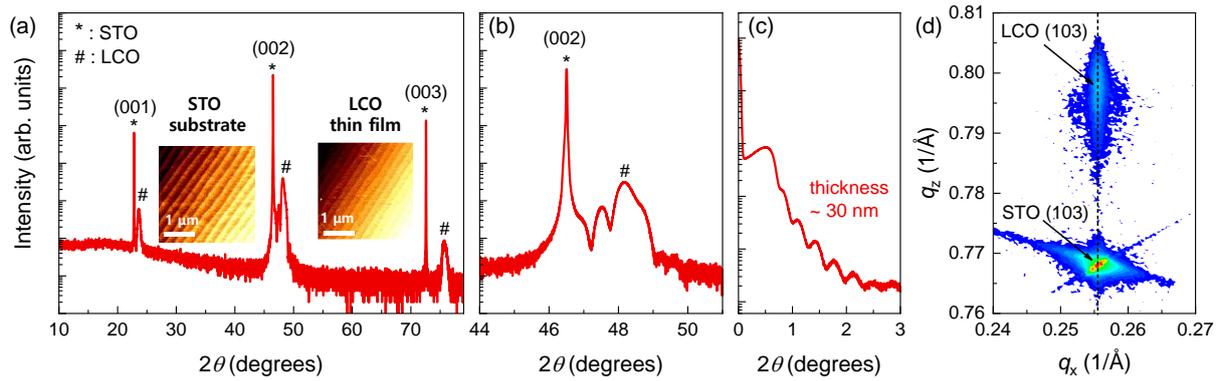

**Figure S1.** Crystal structures of epitaxial LCO thin film on STO substrate. a) XRD $\theta$-$2\theta$ scans for epitaxial LCO thin films (#) grown on STO substrates (*). The insets show AFM images of the sample surface before (left) and after (right) LCO film growth. The root-mean-square of surface roughness was typically below 0.4 nm. The scale bar corresponds to 1 μm. b) XRD $\theta$-$2\theta$ scans near the (002) Bragg plane of the epitaxial LCO thin film on STO substrates. c) The thickness of LCO thin film on STO substrate is estimated by XRR measurement, which value is around 30 nm. d, XRD reciprocal space maps (RSM) of the LCO thin films around the (103) Bragg reflection of the STO substrate. The film is fully strained on the STO substrate.

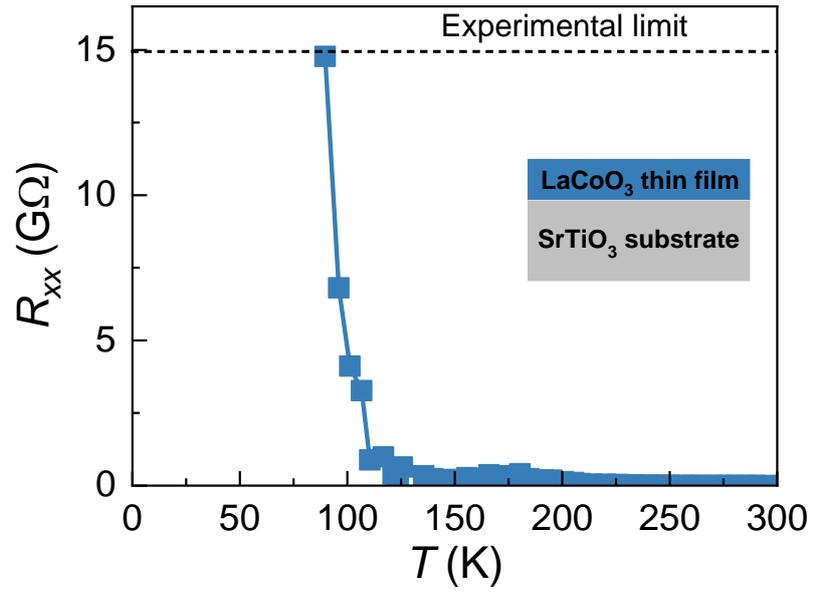

**Figure S2.** *T*-dependent longitudinal resistance $R_{xx}$ of LCO thin films fabricated on STO substrate. The longitudinal resistance $R_{xx}$ as a function of *T* shows the highly insulating behavior of the film. The unit of *y*-axis is $10^9$ Ω.

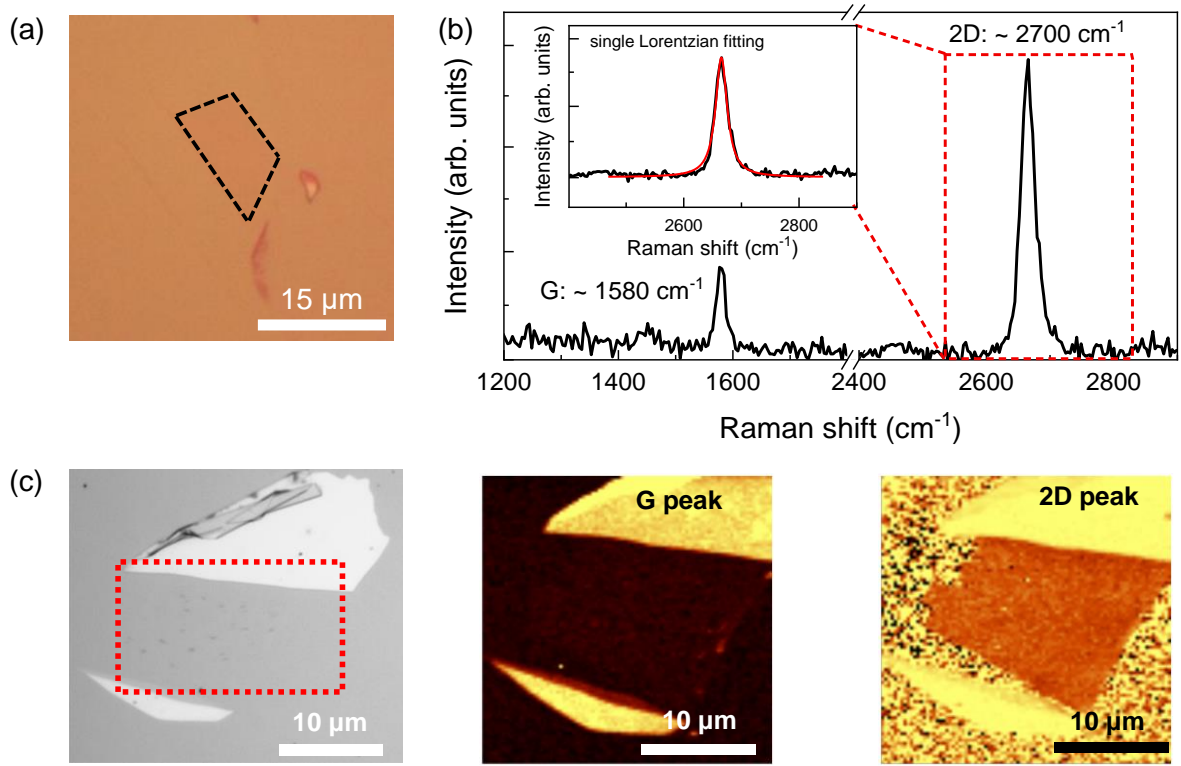

**Figure S3.** The characterization of monolayer graphene flake. a) An optical image of the exfoliated graphene flake on SiO$_2$ substrate. Black dashed lines indicate the shape of the monolayer graphene flake. The scale bar corresponds to 15 μm. b) Raman spectrum of the exfoliated graphene confirming the monolayer thickness. The inset shows that G and 2D peaks are clearly shown and 2D peak is fitted by a single Lorentzian fitting to determine monolayer graphene. c) Raman spectroscopy mapping for the homogeneity of graphene/LCO hybrid heterostructures. The monolayer graphene is homogeneous without any peak shifts of the G (~1,580 cm$^{-1}$) and 2D (~2,700 cm$^{-1}$) peaks, supporting the spin splitting nature observed in the transport experiments.

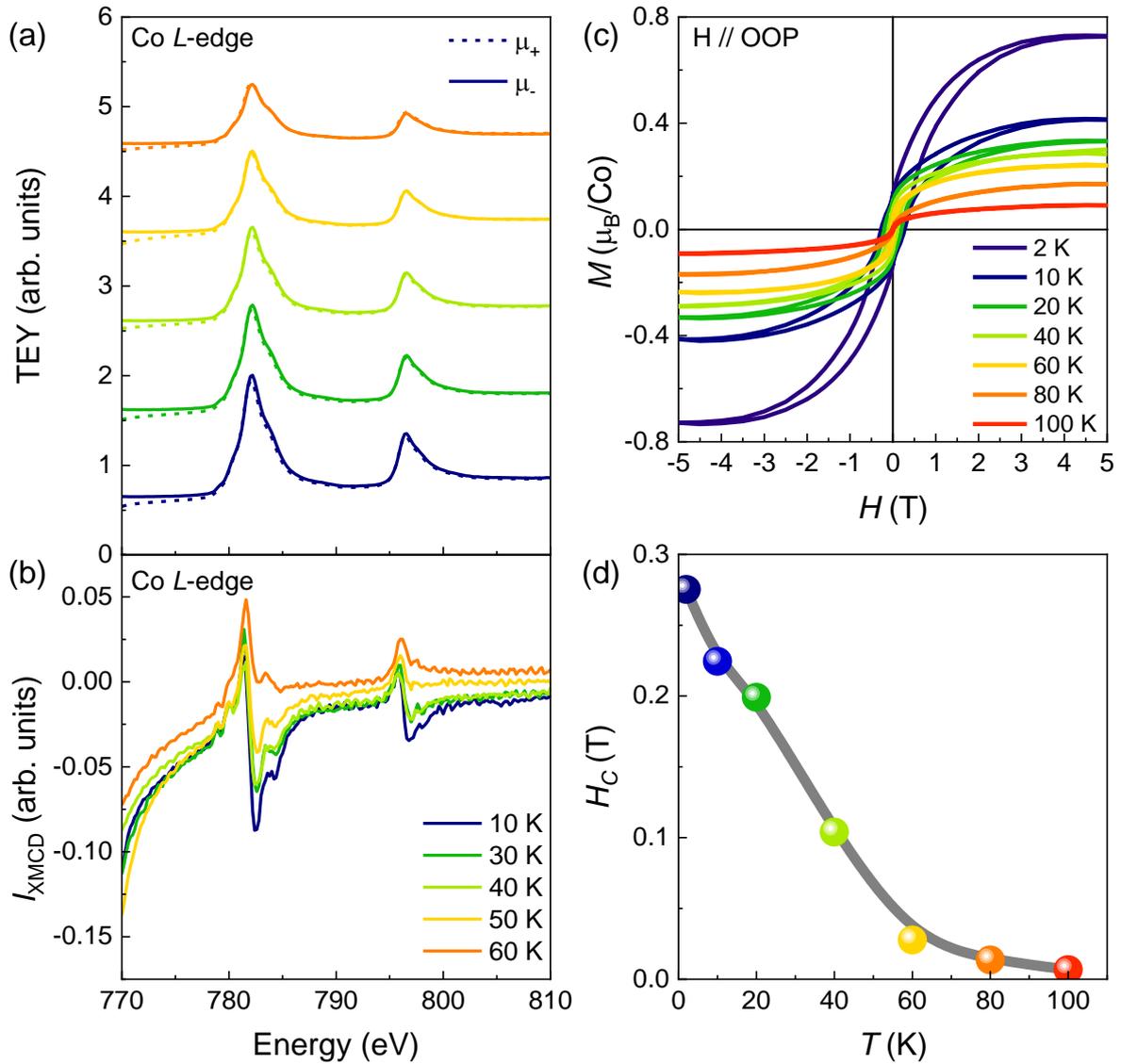

**Figure S4.** Magnetic properties of graphene/LCO hybrid heterostructure along the out-of-plane direction. XAS spectra obtained of the a) Co $L$-edge for both right- and left-handed polarizations of the light with 0.8 T of the magnetic field along the normal direction of the sample surface. b) Acquired XMCD asymmetries spectra of Co $L$-edge demonstrate spin-polarized evidence is significant below 60 K at Co $L$-edge. c) $M(H)$ curves of LCO thin film fabricated on STO substrate were obtained up to 100 K. d) $T$-dependent coercive field ($H_c$) shows hysteresis loop opening below 40 K. Element-specific magnetic characterization in graphene/LCO hybrid heterostructure. Above 60 K, both b) $I_{XMCD}$ and d) $H_c$ and signal were identically suppressed.

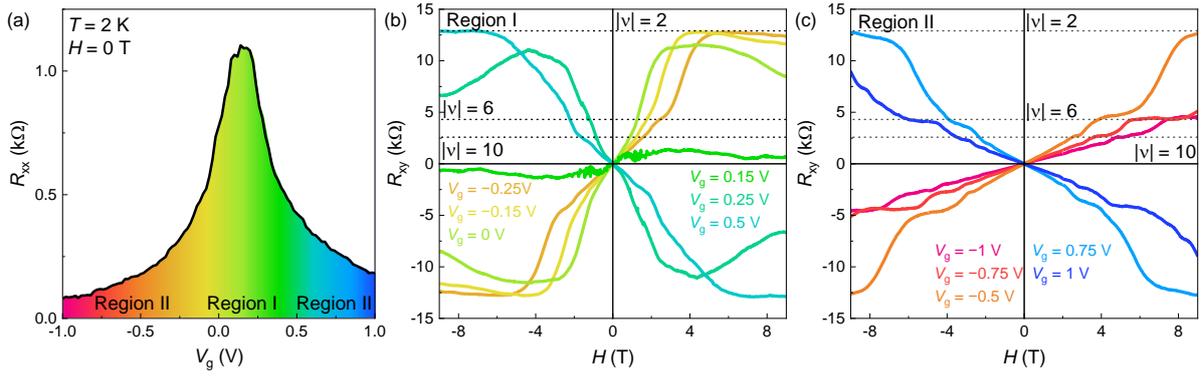

**Figure S5.** Electrical transport characterization of the graphene/LCO hybrid heterostructure. a) Longitudinal resistance $R_{xx}$ as a function of $V_g$ at 2 K and 0 T for the graphene/LCO hybrid heterostructure. The $R_{xx}$ curve reveals a CNP at 0.15 V, and the carrier types are effectively modulated from the hole- (< 0.15 V) and electron-dominant carriers (> 0.15 V). The relatively low and high carrier density areas are separated as Region I (red- and blue-color filled) and Region II (yellow- and green-color filled), respectively. The Hall resistance $R_{xy}$ as a function of the external $H$-field measured at 2 K with various $V_g$ are shown for b) Region I and c) Region II. Horizontal dotted lines represent the filling factors ($|v|$ = 2, 6, and 10).

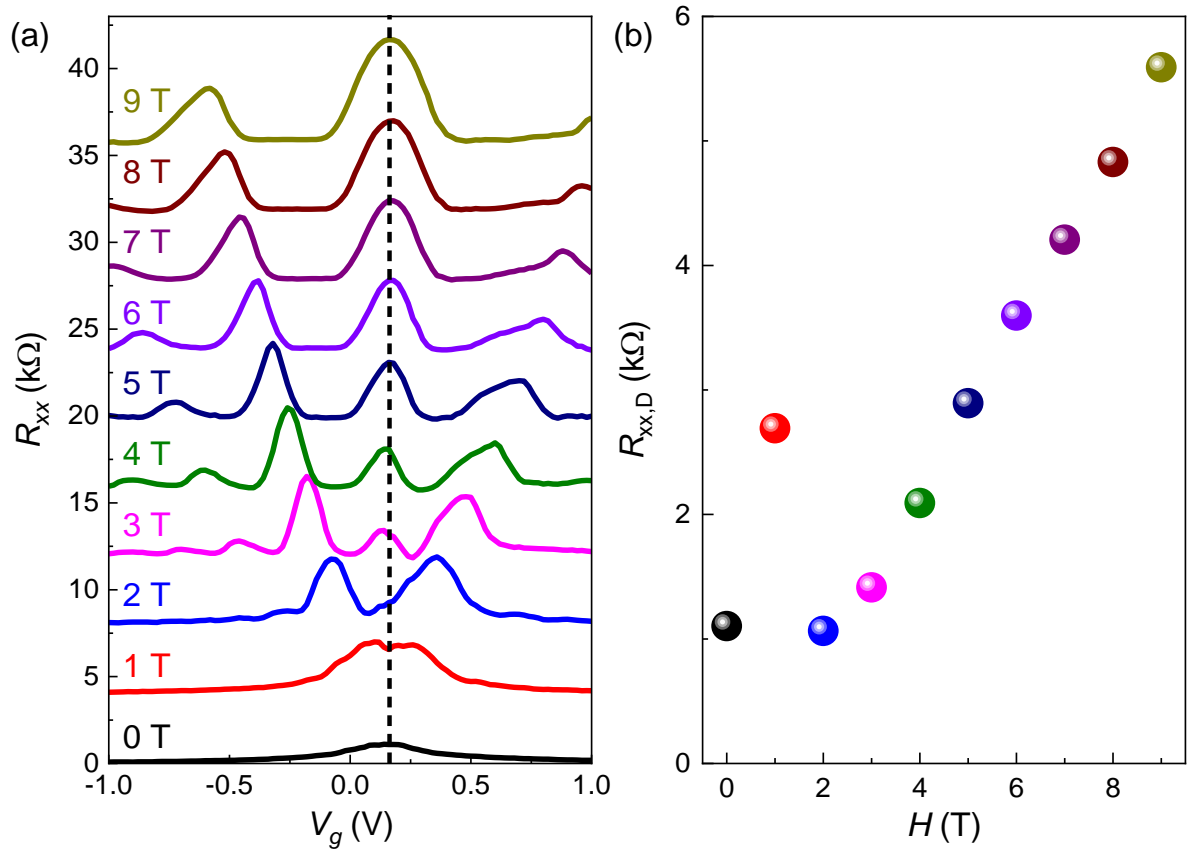

**Figure S6.** Evidence of spin exchange splitting near CNP under varying *H*-field at 2 K in the graphene/LCO hybrid heterostructure. a) Longitudinal resistance $R_{xx}$ as a function of $V_g$ at 2 K and 0 to 9 T for the graphene/LCO hybrid heterostructure. The $R_{xx}$ curves under the different *H*-fields consistently exhibit a CNP near 0.15 V. b) The nonmonotonic behavior of $R_{xx}$ peak at the Dirac point $R_{xx,D}$ as a function of *H*-field.

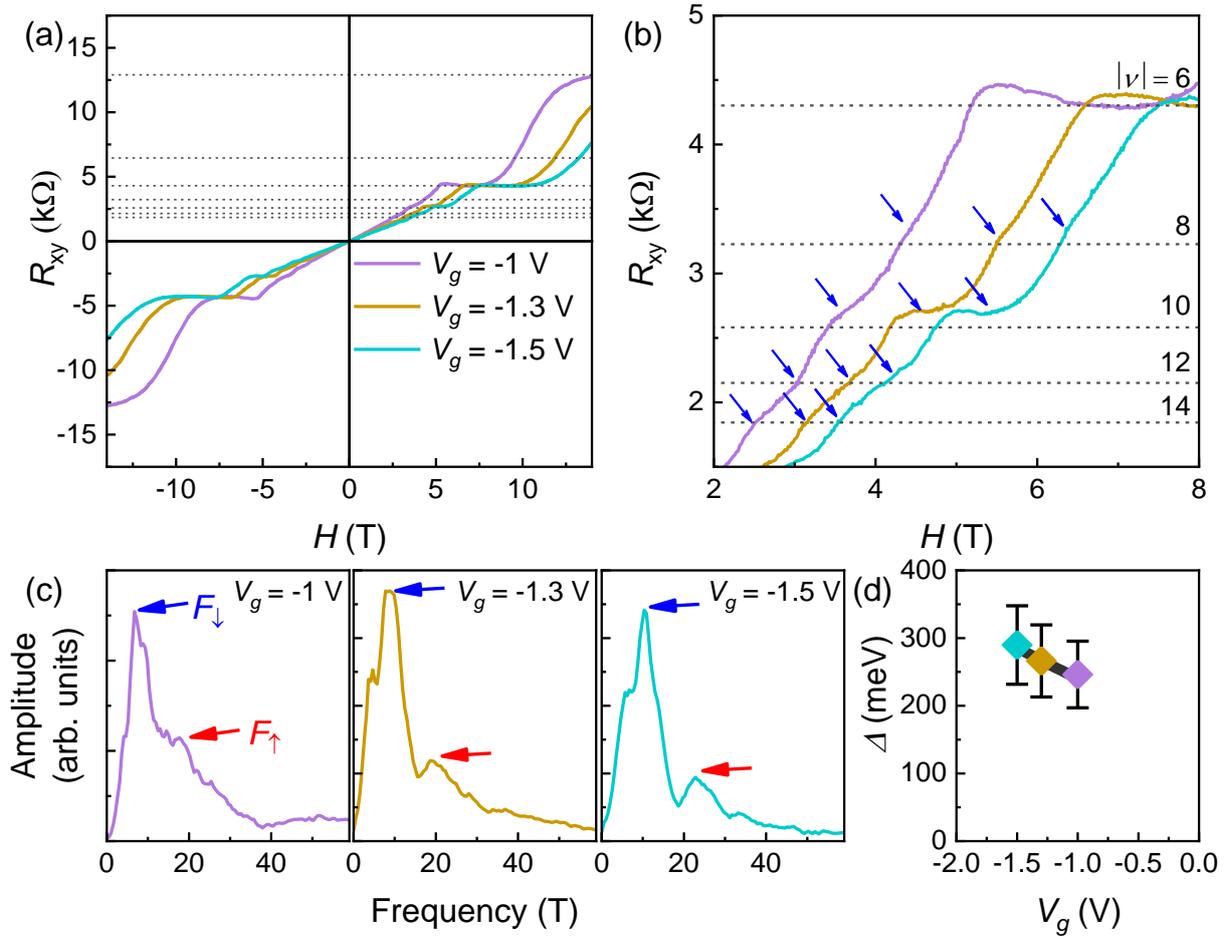

**Figure S7.** The repeated quantum Hall measurements for different graphene/LCO hybrid heterostructure. a) The Hall resistance $R_{xy}$ as a function of the external $H$-field measured at 2 K with various $V_g$ are shown. b) The expanded view of $R_{xy}(H)$ for possible quantized states with spin-symmetry breaking. The dashed lines indicate $|v| = 6, 8, 10, 12$ and $14$. Blue arrows present kinks, which demonstrate spin-symmetry breaking quantized states. c) $V_g$-dependent FFT analysis with distinct two-peak structures shown as features of up- and down-spin. d) $V_g$-dependent $\Delta$ estimated from c at 2 K for various $V_g$.

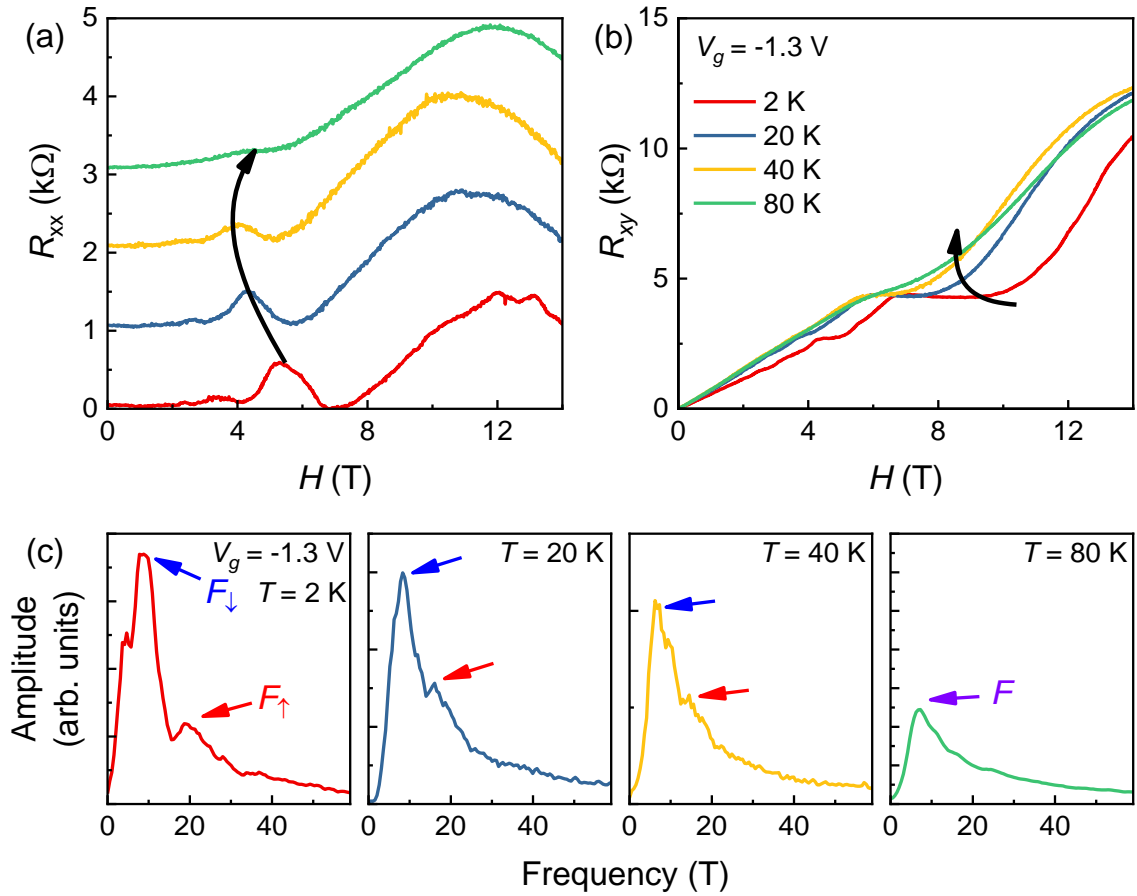

**Figure S8.** Characteristic *T*-dependent behavior of SdH oscillations and quantum Hall plateaus up to 80 K on a different graphene/LCO hybrid heterostructure. a) $R_{xx}(H)$ at $V_g = -1.3$ V at different *T*, vertically shifted (by 1 kΩ) for clarify. The black arrow shows the anomalous shift of the local maxima of $R_{xx}$. b) The Hall resistance $R_{xy}$ as a function of the external *H*-field measured at 2 K with various *T* are shown. c) FFT analysis of $R_{xx}(1/H)$ curves at 2 to 80 K. Two-peak structure below 40 K, in contrast to the single peak above 40 K.

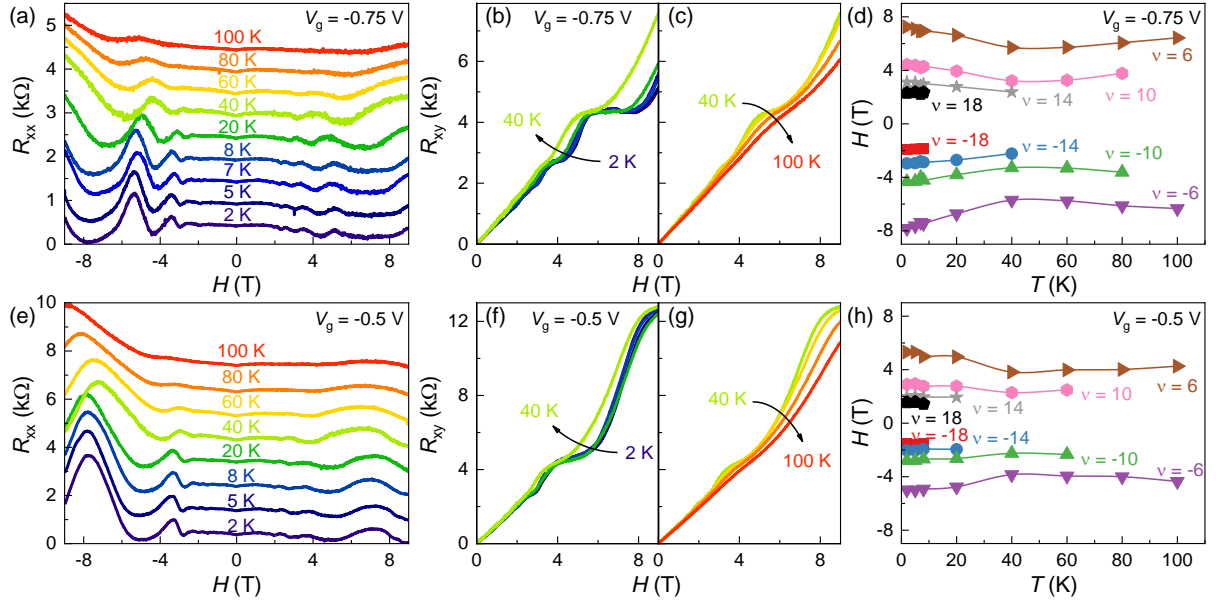

**Figure S9.** Horizontal shift of SdH quantum oscillation for $R_{xx}$ and quantum Hall plateau for $R_{xy}$, respectively by spin splitting in hole doping region. SdH quantum oscillations are horizontally shifted by spin splitting up to 100 K in the hybrid heterostructure. $T$-dependence on $R_{xx}$'s with a) $V_g = -0.75$ V and e) $-0.5$ V are vertically displayed for clarity. Hall resistance $R_{xy}$ also shows same trend below 40 K [b) $V_g = -0.75$ V and f) $-0.5$ V] and above 40 K [c) $V_g = -0.75$ V and g) $-0.5$ V]. d,e, Summary of $T$-dependence on filling factors ($v = $ 6, 10, 14, and 18) with d) $V_g = -0.75$ V and e) $-0.5$ V, respectively.

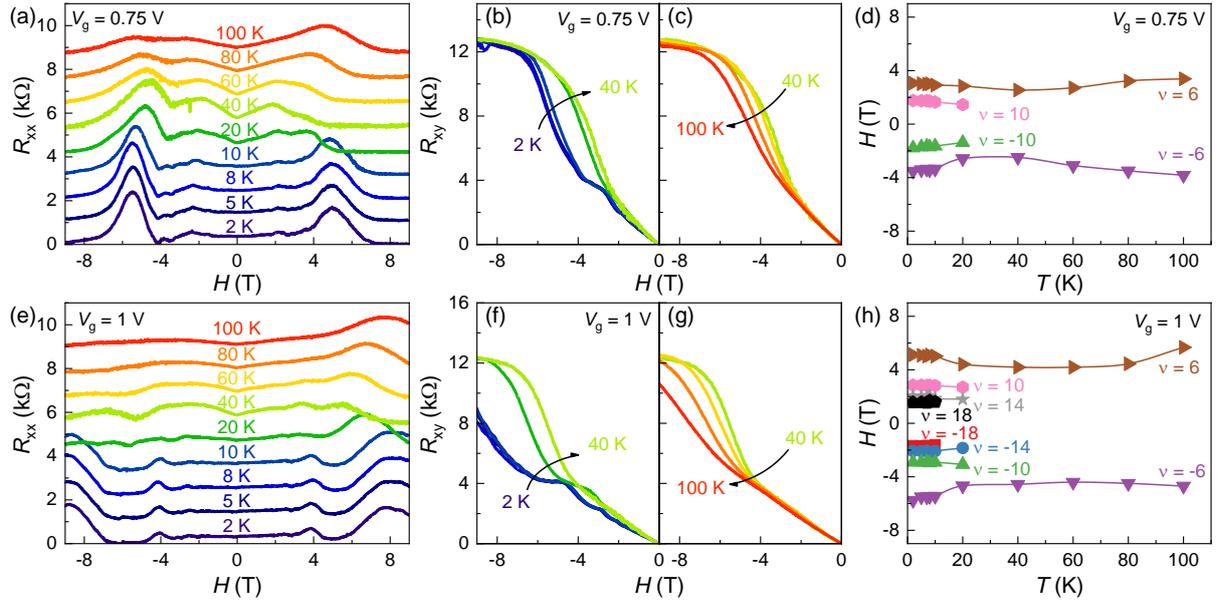

**Figure S10.** Horizontal shift of SdH quantum oscillation for $R_{xx}$ and quantum Hall plateau for $R_{xy}$, respectively by spin splitting in electron doping region. SdH quantum oscillations are horizontally shifted by spin splitting up to 100 K in the hybrid heterostructure. $T$-dependence on $R_{xx}$'s with a) $V_g = 0.75$ V and e) 1 V are vertically displayed for clarify. Hall resistance $R_{xy}$ also shows same trend below 40 K [b) $V_g = 0.75$ V and f) 1 V] and above 40 K [c) $V_g = 0.75$ V and g) 1 V]. Summarizing $T$-dependent filling factors ($\nu = 6, 10, 14,$ and 18) with d) $V_g = 0.75$ V and e) 1 V, respectively.

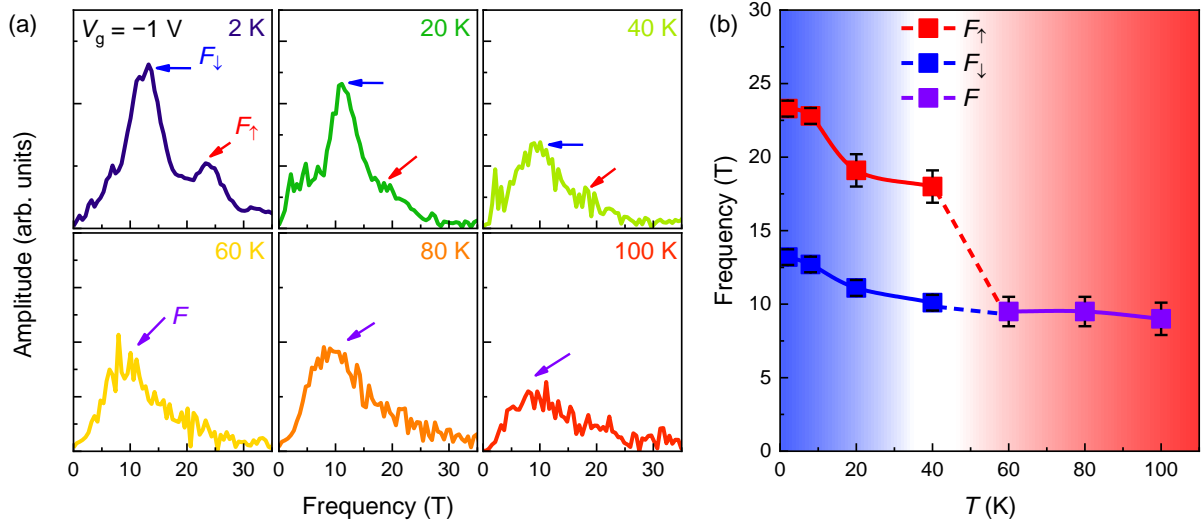

**Figure S11.** Two-peak structure below 40 K, in contrast to the single peak feature above 40 K. a) FFT analysis of $R_{xx}(1/H)$ curves at 2 to 100 K. FFT of SdH oscillation presents Fermi surface at various $T$s, demonstrating that distinct two peaks, which is evidence for the existence of two Fermi surfaces below 40 K with $V_g = -1$ V. Red and blue arrows show Fermi surface of up- and down-spin, respectively, at 2 to 40 K. Purple arrow at 60 to 100 K shows merged Fermi surface without spin splitting. b) $T$-dependence on comparison of plotted Fermi surfaces at $V_g = -1$ V.

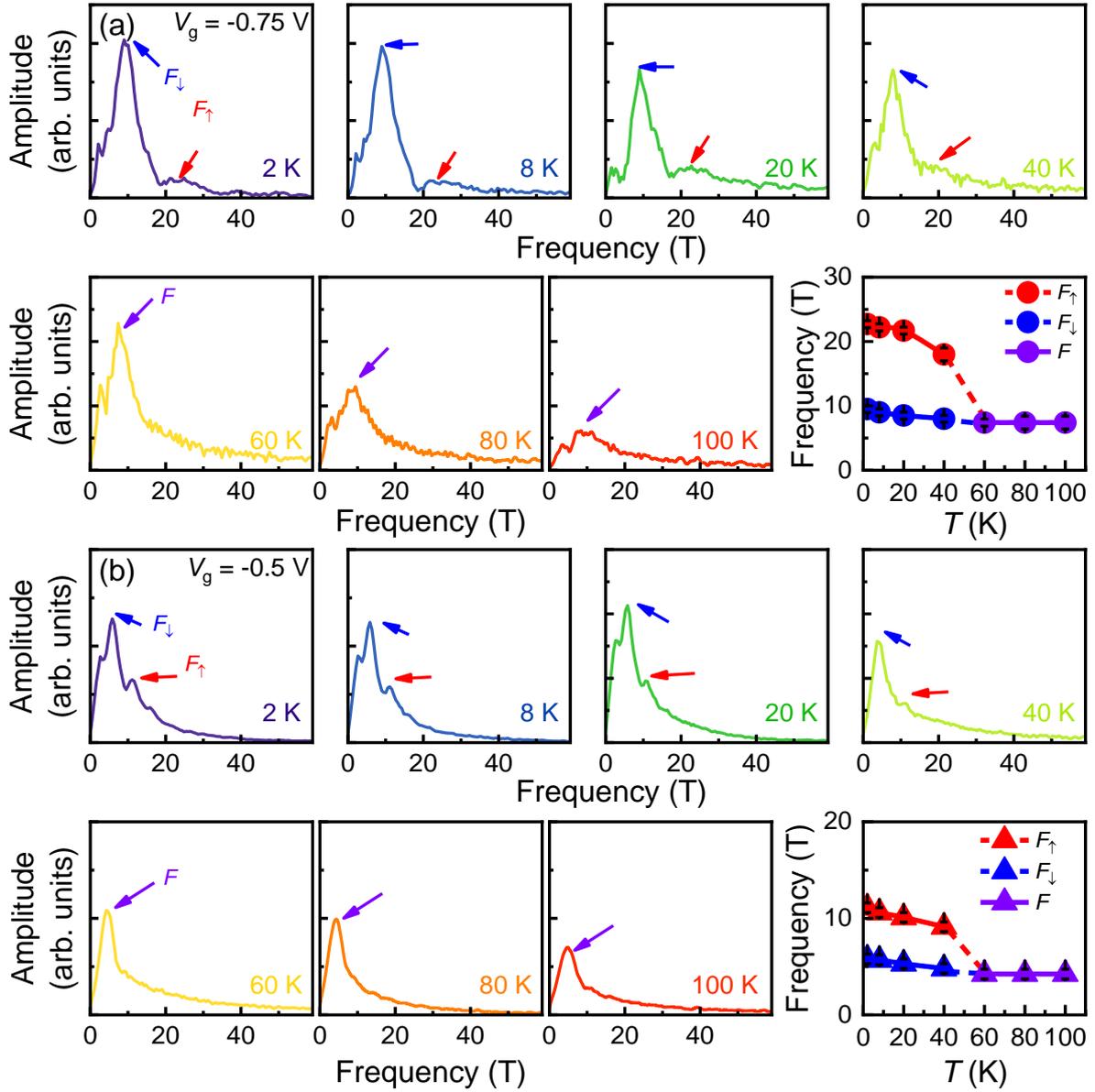

**Figure S12.** Summary of Fermi surface separation in hole doping region. FFT of SdH oscillation presents Fermi surface at various *T*s, demonstrating distinct two peaks, which is evidence for the existence of two Fermi surfaces below 40 K at a) $V_g = -0.75$ V and b) $V_g = -0.5$ V. In contrast, only one Fermi surface is shown above 40 K, since up-and down-spin bands are merged.

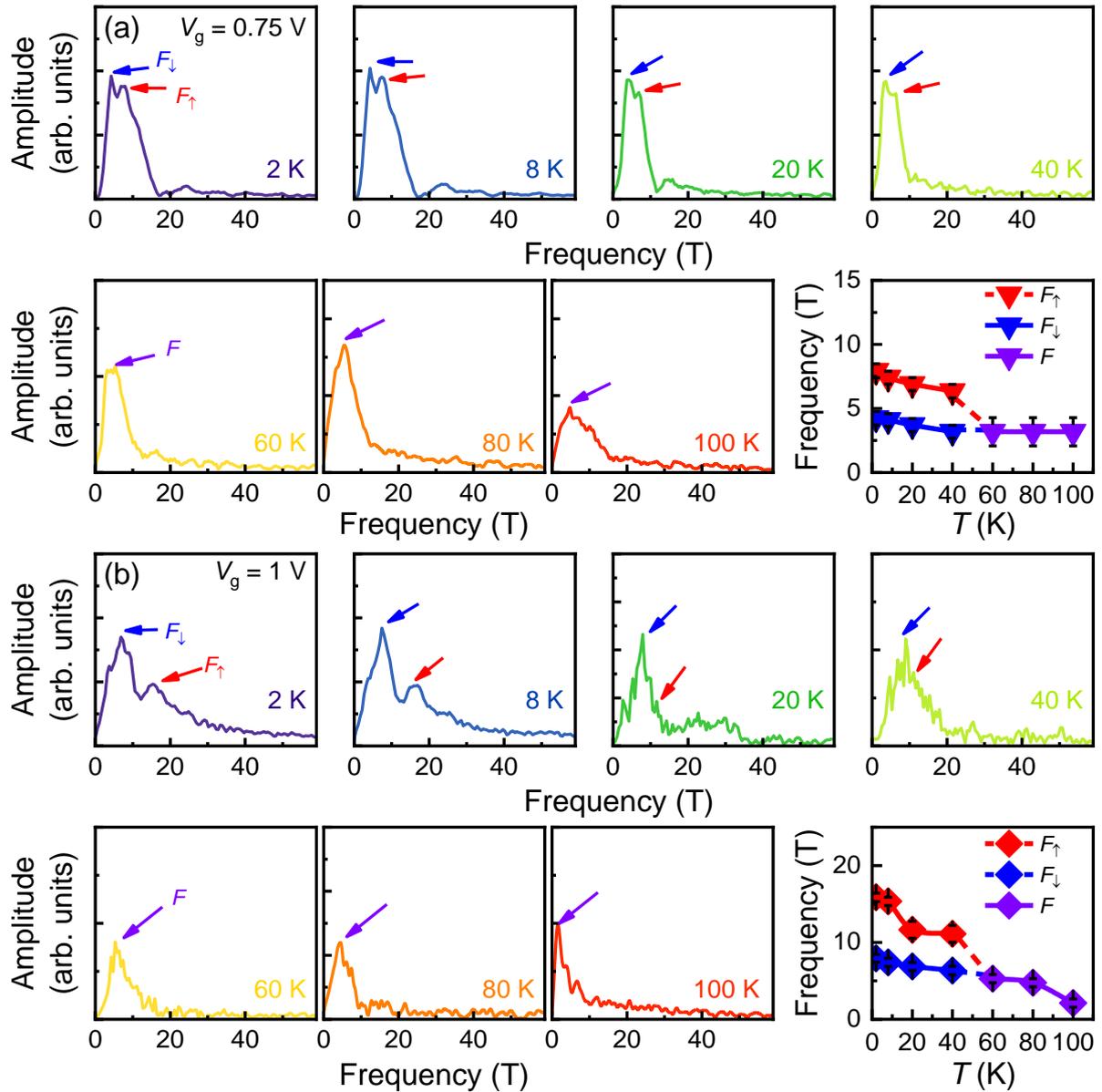

**Figure S13.** Summary of Fermi surface separation in electron doping region. FFT of SdH oscillation presents Fermi surface at various $T$s demonstrating distinct two peaks, which is evidence for the existence of two Fermi surfaces below 40 K at a) $V_g$ = 0.75 V and b) $V_g$ = 1 V. In contrast, only one Fermi surface is shown above 40 K, since up-and down-spin bands are merged.

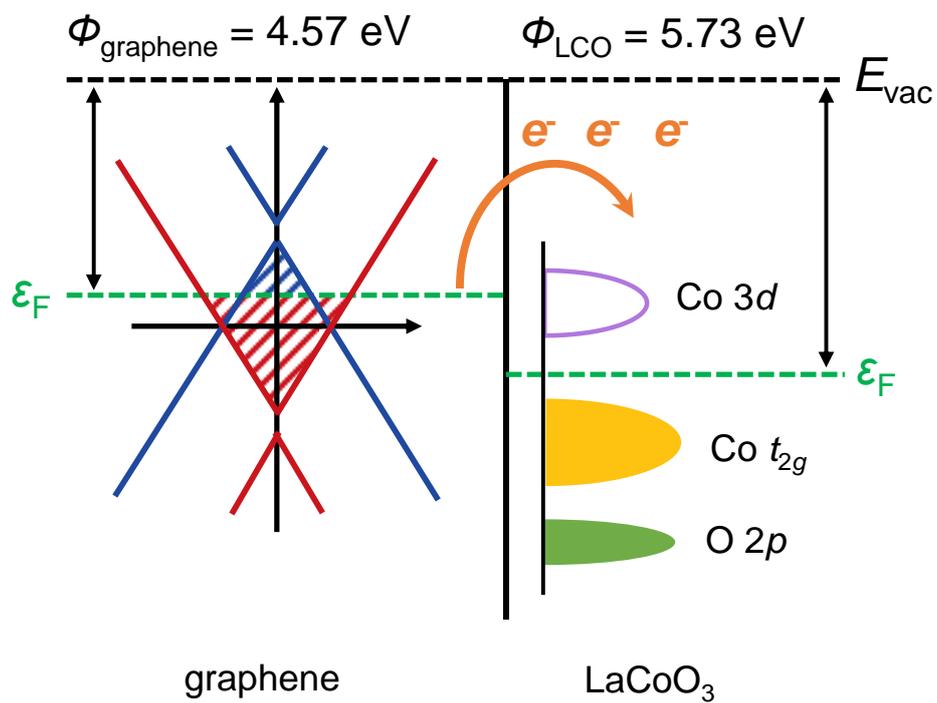

**Figure S14.** Schematic band diagram of interfacial charge transfer between graphene and LCO via the imbalance of work function ($\Phi$).

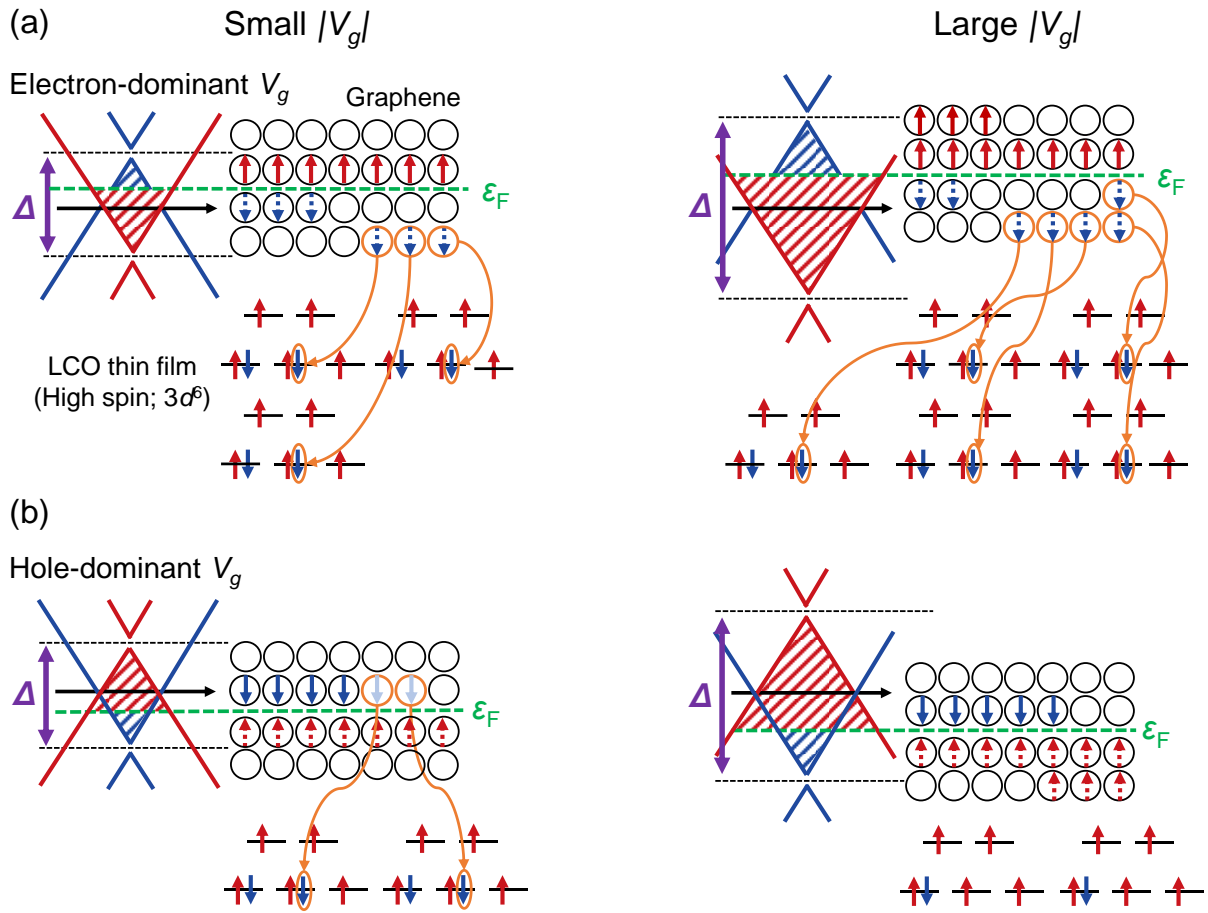

**Figure S15.** Mechanism of interfacial charge transfer. Schematic illustration of interfacial charge transfer within spin-symmetry-broken band dispersions of graphene heterostructured with LCO for a) electron-dominant and b) hole-dominant $V_g$. The red and blue lines represent the linear dispersions near the Dirac point corresponding to the up- and down-spins, respectively. Up- and down-spin carriers are shown as red and blue arrows, respectively, whereas solid and dashed arrows indicate electron- and hole-dominant carriers, respectively. When and only when we consider charge transfer (orange circles), both up- and down-spin Fermi surfaces and $\Delta$ can increase simultaneously with increasing $|V_g|$.